\documentclass[useAMS,usegraphicx,usenatbib]{mn2e}

\newif\ifrefereestyle

\usepackage{multirow}
\usepackage{amssymb}

\def\degrees{^\circ}

\def\arcsec{^{\prime\prime}}
\def\kms{$\mathrm{km\;s^{-1}}$}

\def\pc{{\rm pc}}
   
\newcommand{\re}{\mbox{$R_{\rm eff}$}}
\newcommand{\Msun}{\mbox{$\rm M_{\odot}$}}
\newcommand{\Lsun}{\mbox{$\rm L_{\odot}$}}

\def\etal{{et al.\ }}
\def\eg{{\it e.g.\ }}

\def\ie{{\it i.e.\ }}
\def\cf{{\it cf.\ }}

\def\spose#1{\hbox to 0pt{#1\hss}}

\newcommand{\apprla}{\mathrel{\raise2pt\hbox{\rlap{\hbox{\lower6pt\hbox{$\sim$}}}\hbox{$<$}}}}
\newcommand{\apprga}{\mathrel{\raise2pt\hbox{\rlap{\hbox{\lower6pt\hbox{$\sim$}}}\hbox{$>$}}}}

\newcommand{\Msig}{\mbox{$\rm M_{\bullet}-\sigma$}}
\newcommand{\Mbh}{\mbox{$\rm M_{\bullet}$}}

\newcommand{\Mnc}{\mbox{$\rm M_{\rm NC}$}}
\newcommand{\Mncsig}{\mbox{$\rm M_{\rm NC}-\sigma$}}
\newcommand{\Vrms}{\mbox{$\rm V_{rms}$}}

\title[Dynamical models of the nuclear star cluster in NGC~4244]{Three-integral
multi-component dynamical models and simulations of the nuclear star cluster in
NGC~4244}

\author[De Lorenzi \etal]{F. De
Lorenzi$^1$\thanks{E-mail: flavio.delorenzi@gmx.ch}, Markus Hartmann$^2$, Victor
P. Debattista$^{3,4}$\thanks{RCUK Academic Fellow}, A.C. Seth$^5$, O.
Gerhard$^6$ \\ $^1$ Z\"urcher Hochschule f\"ur Angewandte Wissenschaften,
Technikumstrasse 9, CH-8401 Winterthur, Switzerland \\ $^2$ Astronomisches
Rechen-Institut, Zentrum f\"ur Astronomie der Universit\"at Heidelberg (ZAH),
M\"onchhofstr. 12-14, 69120 Heidelberg, Germany \\ $^3$ Jeremiah Horrocks
Institute, University of Central Lancashire, Preston, PR1 2HE, UK \\ $^4$
Visiting Lecturer, Department of Physics, University of Malta, Tal-Qroqq
Street, Msida, MSD 2080, Malta \\ $^5$ University of Utah, Salt Lake City, UT
USA \\ $^6$ Max-Planck-Institut f\"ur Ex. Physik, Giessenbachstra{\ss}e, D-85741
Garching, Germany}

\begin{document} 

%
\date{Draft, last update \today}
\pagerange{\pageref{firstpage}--\pageref{lastpage}} \pubyear{----}
\maketitle

\label{firstpage}

\begin{abstract}

  Adaptive optics observations of the flattened nuclear star cluster in the
  nearby edge-on spiral galaxy NGC~4244 using the Gemini Near-Infrared Integral
  Field Spectrograph (NIFS) have revealed clear rotation. Using these
  kinematics plus 2MASS photometry we construct a series of axisymmetric
  two-component particle dynamical models with our improved version of {\sc
  nmagic}, a flexible $\chi^2$-made-to-measure code. The models consist of a
  nuclear cluster disc embedded within a spheroidal particle population.  We
  find a mass for the nuclear star cluster of
  $\mathrm{M}=1.6^{+0.5}_{-0.2}\times 10^7 \Msun$ within $\sim 42.4$~pc
  ($2\arcsec$).  We also explore the presence of an intermediate mass black
  hole and show that models with a black hole as massive as $\Mbh = 5.0 \times
  10^5~\Msun$ are consistent with the available data.  Regardless of whether a
  black hole is present or not, the nuclear cluster is vertically anisotropic
  ($\beta_z < 0$), as was found with earlier two-integral models.
  We then use the models as initial conditions for $N$-body
  simulations.  These simulations show that the nuclear star cluster
  is stable against non-axisymmetric perturbations.  We also explore
  the effect of the nuclear cluster accreting star clusters at various
  inclinations.  Accretion of a star cluster with mass $13\%$ that of
  the nuclear cluster is already enough to destroy the vertical
  anisotropy, regardless of orbital inclination.

\end{abstract}   


\section{Introduction}   
\label{sec:intro}

Studies of the centres of galaxies across the Hubble sequence have
shown that they frequently host central massive objects such as
massive nuclear star clusters (NCs) and supermassive black holes
(SMBHs).  NCs are present in roughly 75\% of low and intermediate
luminosity disc and elliptical galaxies \citep{boe_etal_02,
  cot_etal_acsvcs8_06}.  These NCs are intrinsically very luminous,
with typical $M_I \sim -12$, and sizes similar to globular clusters
\citep[$r_{eff} \sim 5$pc;][]{boker04a}.  

Two hypotheses have been offered to explain NC formation.  One
scenario envisages NCs forming in situ out of gas cooling onto the
centre \citep{milosa_04, bek_etal_06, bekki_07}.  Alternatively, NCs
may form from star clusters merging at the centres of galaxies 
  \citep{tre_etal_75, lot_etal_01, cap-dol_mio_08, agarwal+2011,
    antonini+2012a, antonini2012b}. Which hypothesis is correct
determines whether NC growth is limited by the supply of star clusters
from the host galaxy \citep{antonini2012b} or regulated by
feedback from in situ star formation \citep{mcl_etal_06}.

The assembly history of NCs can be constrained from their morphology,
stellar populations and kinematics.  In late-type spirals, NCs have
been found to consist of multiple stellar populations, typically a
young population ($<100$ Myr), and a dominant population older than 1
Gyr \citep{dav_cou_02, sch_etal_03, rossa+06, wal_etal_06}.  The {\it
  Hubble Space Telescope} has revealed that the NCs of edge-on
galaxies host multiple stellar populations associated with different
morphological components \citep{seth06}.  These NCs consist of young
blue nuclear cluster discs (NCD) and older nuclear cluster spheroids
(NCS).  Optical spectra of the NC in the edge-on Scd galaxy NGC~4244
($i\approx90\degrees$), the nearest galaxy in the sample of
\citet{seth06} \citep[D=4.37~Mpc;][]{seth05a}, indicates the presence
of multiple stellar populations, while near infrared spectroscopy
showed that the NC is rapidly rotating \citep{seth+08b}. Using
$N$-body simulations \citet[][hereafter H11]{hart_etal_11} showed that
the NC in NGC 4244 cannot have assembled more than half its mass via
the accretion of star clusters.  

NCs exhibit several scaling relations.  The luminosity of NCs
correlates with that of their host galaxy \citep{boe_etal_02,
  cot_etal_acsvcs8_06, erwin+gadotti10}.  A number of studies also
found that their mass, \Mnc, correlates with the velocity dispersion
of the host bulge, the \Mncsig\ relation \citep{fer_etal_06,
  weh_har_06, rossa+06}.  Early work found that this \Mncsig\ relation
is parallel to the \Msig\ relation of SMBHs \citep{geb_etal_00,
  fer_mer_00}, with NCs being about $10\times$ more massive, at the
same $\sigma$, as SMBHs.  However recent work has questioned how
comparable NCs and SMBHs are.  \citet{erwin_etal_2012} find that NCs
and SMBHs follow different relations, with SMBH masses correlated with
properties of the bulge, while NCs seem to correlate better with
properties of the entire host galaxy. Instead, both \citet{leigh+12}
and \citet{scott_graham12} show that there is an $\Mnc-\sigma$
relation but with a significantly different slope than for SMBHs.
It is not clear at present whether these differences are intrinsic to NC and
    SMBH growth or whether they are due to the fact that the scaling relations
    depend on Hubble type.  In particular, some recent studies have suggested
    that SMBHs and NCs  in late-type galaxies do not follow the same scaling
    relations as in early-types \citep{greene+2010, erwin+gadotti12}.  
    Some galaxies host both a NC and a SMBH
\citep{seth08, graham+spitler_09b}. The relative
  properties of NCs and SMBHs in such galaxies could constrain the
  relationship between these objects.  For instance, by constructing
  an $(\Mbh+\Mnc)-\sigma$ relation which includes the mass of both the
  NC and of the SMBH \citep{graham_etal_11}, \citet{graham12} found a
  flatter relation than the \Msig\ relation.  But the small existing
  sample of objects with known NCs and SMBHs is currently too small to
  obtain a clear picture \citep[e.g.][]{neumayer_etal_12}.  Progress
  in determining whether NCs and SMBHs are related therefore requires
  improving the statistics of such measurements.  Moreover, a better
  understanding of the mass assembly of NCs in late-type galaxies is
  vital.

It is generally thought that AGN feedback is responsible for the
\Msig\ relation \citep[e.g.,][]{sil_ree_98, aking_03,
    mur_etal_05, dim_etal_05, saz_etal_05, spr_etal_05b,
    joh_etal_09}, but scenarios where this relation arises because
the galaxy regulates SMBH growth \citep[e.g.][]{bur_sil_01,
  kaz_etal_05, mir_kol_05} or purely indirectly by the
  hierarchical assembly through galaxy merging \citep{hae_kau_00,
  ada_etal_01, ada_etal_03, jahnke_maccio10} have also been proposed.
  If gas inflow plays a more important role in the growth of NCs
  then this opens the possibility that some form of feedback drives
  the scaling relations in both SMBHs and NCs
  \citep[e.g.][]{mcl_etal_06}.

To help shed light on the formation of NCs in late-type galaxies, in
this paper we study the NC in the nearby Sc galaxy NGC 4244. H11
  modelled this NC using two-integral {\sc jam} models
  \citep{cappellari08}, obtaining a mass of $(1.1\pm 0.2) \times
  10^7~\Msun$.  In this paper we build three-integral particle models
  of the same NC and use them as initial conditions for $N$-body
  simulations to explore its sensitivity to star cluster accretion.
The outline of this paper is as follows.  Section~\ref{sec:obsdata}
describes the observational data and how they are used in the
dynamical modelling.  Our modelling method, the $\chi^2$-M2M code {\sc
  nmagic} based on \citet{delo+07, delo+08, delo+09}, is described in
Section~\ref{sec:methods} including additional code development.  We
construct various axisymmetric particle models of the nuclear region
of NGC~4244 in Section~\ref{sec:models} using this improved code.  The
models consist of a NCD and NCS having separate mass-to-light ($M/L$)
ratios.  Using the best model as initial conditions for $N$-body
simulations, we explore the evolution of the NC in
Section~\ref{sec:nbody}.  Section~\ref{sec:conclusions} discusses our
results in the context of NC formation.


\section{Observational Data}
\label{sec:obsdata}

We begin by describing the photometry and how these data are
deprojected to obtain a three dimensional luminosity density. After
this, the integral-field kinematic data are presented. We adopt a
distance to NGC 4244 of $4.37\; \rm Mpc$ \citep{seth+05b}.  At this
assumed distance, 1$\arcsec$ corresponds to 21~pc.

\subsection{Photometry}
\label{sec:photo}

Here we give a brief summary of the photometric data and its model
representation, both described in detail in \citet{seth05a, seth+08b}.

The photometry consists of $K$-band data either from 2MASS or from the
NIFS observations of \citet{seth+08b}. The main disc (MD) of NGC 4244
hosts at its centre a NC which is composed of a NCD and an oblate NCS.
The $K$-band mass-to-light ratio $M/L_K$ is estimated to be $0.5-0.75$
for the galaxy as a whole (from integrated colours taken from LEDA
[$B-V=0.4-0.6$] combined with $M/L$ from \citet{bell+03}).  The
$M/L_K$ of the NCD is in the range $0.1-0.25$ based on
\citet{bruzual_charlot03} models applied to the optical spectroscopy
of \citet{seth06}, which agrees well with the fitted luminosities for
the disc in {\it HST}/ACS and NIFS bands. The NCS stellar populations
are poorly constrained and a $M/L_K$ between $0.5$ and $1.2$ is
likely.

The surface brightness of NGC~4244 is decomposed into an axisymmetric
three component model. The luminosity distributions of the MD and the
NCD are modelled as projected edge-on exponential discs
\citep[\eg][]{vanderkruit81}:
\begin{equation}
\Sigma(x,z) = \Sigma_0 \left ( \frac{x}{h_r} \right ) K_1(x/h_r)\,
\mathrm{sech}( z/z_0)^2, \label{eqn:SBDisk}
\end{equation}
where $\Sigma_0$ and $h_r$ are constants, and $K_1$ is the modified
Bessel function. The corresponding model parameters for the MD and NCD
models are taken from \citet{seth05a} and \citet{seth+08b},
respectively.

On the other hand the NCS is represented using a \citet{sersic68}
profile:
\begin{equation}
I(x,z) = I_e \exp(-b_n((R/R_e)^{1/n}-1)),
\end{equation}
where $I_e$ is the surface brightness at the effective radius $R_e$,
$b_n \simeq 1.992 n-0.3271$ and $R=\sqrt{x^2+(z/q)^2}$ is the
elliptical radius, with flattening $q$. Best fit parameters have been
obtained from \citet{seth+08b}.

All models were convolved with a Gaussian point spread function (PSF)
of $0\farcs23$ FWHM during the fitting process, \cf \citet{seth+08b}.
The best fit parameters are summarised in Table~\ref{tab:photo}. The
left panel of Figure~\ref{fig:photometry}, which presents the NCS
model, shows its surface brightness, $\mu_K$, and ellipticity,
$\epsilon=1-q$, profiles.
\begin{centering}
\begin{table*}
\vbox{\hfil
\begin{tabular}{ccccccccc}
  \hline
  \\[-1.7ex]
  \multirow{2}{*}{\textsc{Comp.}} 
   & $\Sigma_0$ & $\rho_0$ & $h_r$ & $z_0$ & 
  $q$ & $I_e$ & $R_e$ & $n$ \\
   & [$L_{\odot}/{\rm pc^2}$] & [$L_{\odot}/{\rm pc^3}$] & [${\rm pc}$] & [${\rm pc}$] &  & [$L_{\odot}/{\rm pc^2}$] & [${\rm pc}$] &  \\
  \\[-1.7ex] 
  \hline
  \hline
  MD & $598$  & $0.167$ & $1783$ & $469$ &  & & &  \\ 
  NCD & $1.41\times 10^5$  & $2.08\times 10^4$ & $3.39$ & $1.19$ &  & & &  \\ 
  NCS & & & & & $0.81$ & $8.73\times 10^3$ & $10.86$ & $1.68$ \\ 
  \hline
\end{tabular}
\hfil}
\caption{Best-fit parameters  of the photometric three component model 
  taken from \citet{seth05a,seth+08b}. The model consists of an exponential
  main disc (MD), an exponential nuclear cluster disc (NCD) and a S\'ersic
  nuclear cluster spheroid (NCS).} 
\label{tab:photo} 
\end{table*}
\end{centering}

\subsubsection{Deprojection}
\label{sec:depro}

To compute the three dimensional luminosity distribution, each
component of the surface brightness model is deprojected individually.
The edge-on deprojection of an axisymmetric system is unique
\citep{rybicki87}. The surface-brightness profile of
Equation~\ref{eqn:SBDisk} corresponds to an exponential disc, so the
deprojection is readily given by \citep[\eg][]{vanderkruit81}:
\begin{equation}
\rho(R,z) = \rho_0 \exp(-R/h_r)\,\mathrm{sech}( z/z_0)^2, 
\end{equation}
where $\rho_0= \Sigma_0/(2 h_r)$, and $\Sigma_0$, $h_r$ and $z_0$ are
as in Equation~\ref{eqn:SBDisk}. Their values are given in Table
\ref{tab:photo}.

\begin{figure}
\centering 
\includegraphics[angle=-90.0,width=0.49\hsize]{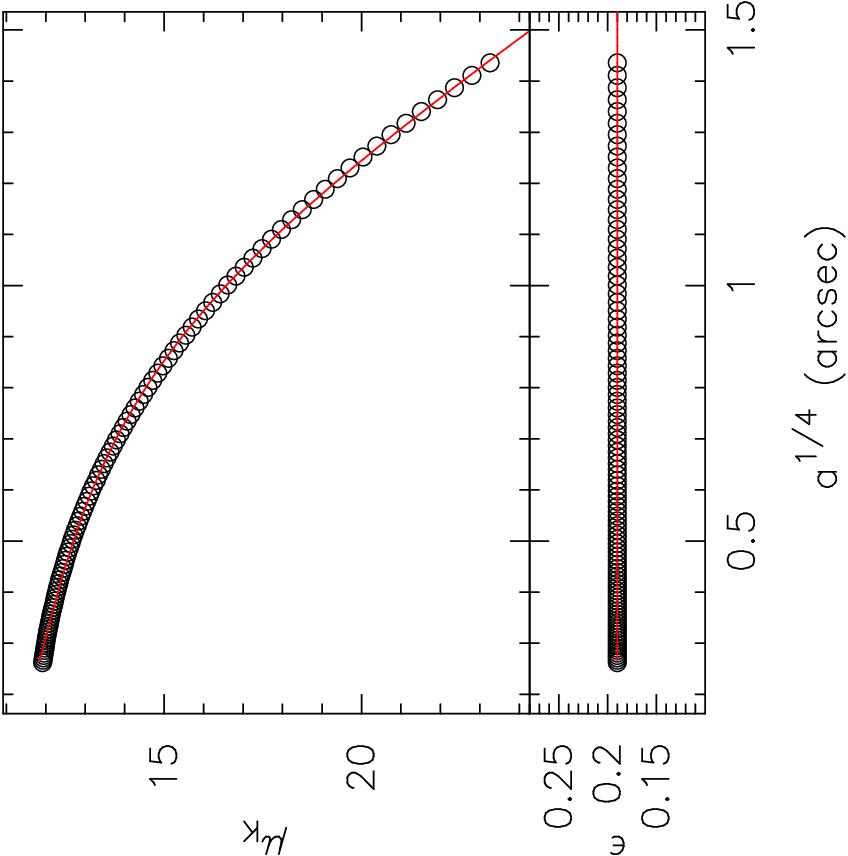}
\hskip0.1truecm
\includegraphics[angle=-90.0,width=0.46\hsize]{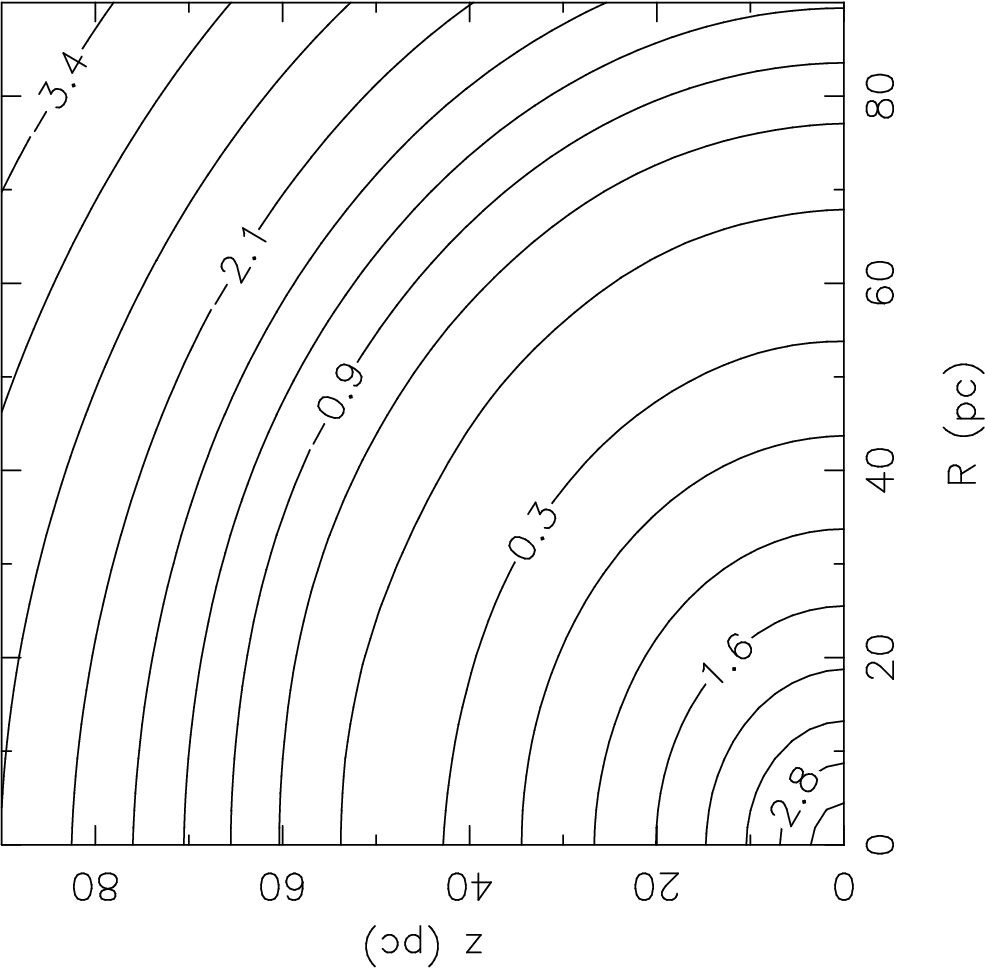}
\vskip0.2truecm
\caption[]{Left: Comparison of the NCS S\'{e}rsic photometry with the
reprojected three dimensional luminosity model. The data points
correspond to the NCS S\'{e}rsic profile, which was derived via a
morphological fit. The solid line shows the model projected onto the
sky plane. The upper panel shows the surface brightness, $\mu_K$, and
the lower one shows the ellipticity, $\epsilon=1-q$.  Right:
Iso-density contours of the NCS in the meridional plane. The
luminosity density was obtained by an edge-on deprojection of the NCS
S\'{e}rsic surface brightness. The contour labels are given in units
of $\log_{10} L_\odot$ {\rm pc}$^{-3}$.}
\label{fig:photometry} 
\end{figure}

Unlike the surface brightness profile of an exponential disc, the
S\'{e}rsic profile of the NCS cannot be deprojected in closed form. We
therefore use the program of \cite{magorrian99} to numerically
deproject the surface brightness distribution of the NCS. The program
finds a smooth axisymmetric density distribution consistent with the
surface brightness distribution for the specified inclination angle
(here edge-on, \ie $90 \degrees$), by imposing that the solution
maximises a penalised likelihood.  Because the deprojection is
computed numerically and tabulated on a grid, the reprojected surface
brightness profile may not match the S\'ersic one perfectly.  A
comparison of the NCS S\'{e}rsic photometric model and its edge-on
deprojection reprojected onto the sky plane, seen in the left panels
of Figure~\ref{fig:photometry}, shows that the numerical deprojection
is in fact very reliable. The right panel presents iso-density
contours in the meridional plane of the NCS obtained with
\cite{magorrian99}'s code.

\subsection{Kinematic Data}
\label{sec:kindata}

The integral-field NIFS kinematic data we use in the modelling were
presented in \citet{seth+08b} and consist of velocity, velocity
dispersion and the higher order Gauss-Hermite moments $h_3$ and $h_4$
\citep{vdmarel_franx93, gerhard93}.  The NIFS field-of-view extends to
$\pm 1.5\arcsec$ along each direction, but the usable data is within
$\pm 1.0\arcsec$. In this field-of-view, the positions of the
``spaxels'' within which spectra were taken define a grid of $63\times
71 = 4473$ cells, which serves as the basis grid for Voronoi bins for
which the velocity, velocity dispersions, $h_3$ and $h_4$ are given.


\section{Methods}
\label{sec:methods}

We construct a range of dynamical models for the NC of NGC 4244.
These models consist of a disc and a spheroidal particle population
representing the NCD and NCS respectively. We use an adapted version
of the flexible $\chi^2$-made-to-measure ({\sc M2M}) particle code
{\sc nmagic} described in \citet{delo+07, delo+08}.  This section
describes a few ingredients required to construct dynamical M2M models
and presents further development of {\sc nmagic} compared with our
previous work in \citet{delo+07, delo+08}.

\subsection{Model observables}
\label{sec:obs}

The central luminosity volume density of the NCS is about a million
times larger than that of the main disc \citep{seth+08b}. Thus the NCS
dominates the luminosity distribution in the central region out to the
edge of the observational data, allowing us to neglect the luminosity
distribution of the main disc when computing the photometric
observables. At a distance $R=30$~pc from the centre, the luminosity
density in the equatorial plane of the NCS is still about $100$ times
that of the MD.

We compute separate spherical harmonic coefficients $A_{lm}$ for the
density of the NCD and the NCS. The corresponding errors are inferred
following a Monte-Carlo procedure described in \citet{delo+08}, in
which the $A_{lm}$'s are computed many times from random rotations
about arbitrary axes of a suitable particle realisation. Here, we used
the isotropic particle model generated from the major-axis density
profiles of the NCS and NCD components, described in
Section~\ref{ssec:ics} below. The $A^{\rm NCD}_{lm}$ and $A^{\rm
  NCS}_{lm}$ constrain the photometry of the disc and spheroidal
particle populations, respectively. We use even $A_{lm}$'s up to
$l_{max}=8$ in 40 radial bins, unevenly spaced, for a total of
$N_{Alm}=2000$ photometric constraints. The grid starts at
$r_{min}=0\farcs0001$ ($0.002$~pc) and extends to $r_{max}=2\,\arcsec$
($42.4$~pc).

The NIFS kinematic data (velocity $v$, dispersion $\sigma$, $h_3$ and
$h_4$) are bi-symmetrised by adopting a point-symmetric reflection with
respect to the centre of the galaxy, as described in \citet{delo+09},
followed by a reflection about the major axis. The resulting data
within $\pm 0.7 \arcsec$ are shown in the upper panel of Figure
\ref{fig:nifskin_model}. As kinematic observables we use
luminosity-weighted Gauss-Hermite coefficients $h_1$ up to $h_4$
\citep{gerhard93, vdmarel_franx93, delo+08} and the luminosity itself
(corresponding to $h_0$) within the field-of-view, for a total of
$N_{kin}=365$ kinematic observables ($73$ Voronoi bins times $5$ sets
of Gauss-Hermite coefficients).  These are used to constrain the
particle system as a whole, without distinguishing between NCD and
NCS populations. The particle model is seeing convolved with a
Gaussian PSF having a FWHM of $0\farcs23$ ($5$~pc) by means of the
Monte-Carlo method presented in \citet{delo+08}: When the model
kinematics are computed, each particle is temporarily replaced by
$N_{pp}$ pseudo particles with randomly selected positions having
probabilities given by the PSF. In the present work, we adopted
$N_{pp}=5$.

\subsection{Initial conditions}
\label{ssec:ics}

We set up {\it spherical} initial conditions using the major-axis
density profile of the NCS. Setting $M/L_K=1$, the mass density is
normalised to unit mass and the self-consistent gravitational
potential is computed. Following \citet{gerhard91}, the isotropic
distribution function is computed and used to generate a set of equal
mass particles as in \citet{victor_sell00}.  Finally, 30\% of the
particles are randomly assigned to the "disc" population.  We produced
two realisations of these initial conditions, a low resolution one
with 0.75M particles which allowed us to explore parameter space
quickly and a higher resolution version with 6M particles.

\subsection{Gravitational potential}
\label{ssec:pot}

The {\sc M2M} method works by adjusting the weights of individual
particles while they are evolved along their orbits. For this orbit
integration the gravitational potential of the system is needed. This
section details the methods used to construct the dynamical models
presented in Section~\ref{sec:models} below. In brief, low resolution
models (M1) are built using an {\sc FFT} method for the potential.
Then high resolution models without (M2) and with intermediate
mass black holes (IMBHs) (M3--M8) are built with the potential
computed on a spherical mesh of spherical harmonics
\citep{sellwood03}.

We assume that the mass distributions of the MD, NCD and NCS follow
their luminosity distributions $J_i$ ($i\in \{$MD, NCD, NCS$\}$). For
a constant mass-to-light ratio $M/L_i$, the corresponding mass density
is $\rho_i = (M/L_i) J_i$, with $J_i$ the deprojection given in
Section~\ref{sec:depro}.

The potential in models M1 was obtained on a Cartesian grid.  Each
Cartesian grid consists of $N=128^3$ grid cells.  The NC grid extends
to $\pm 50$~pc along each direction, whereas the MD grid extends to
$\pm 3000$~pc. This allows us to resolve the scale-heights of the MD
and the NCD and the half-mass radius of the NCS. The potential on each
grid is calculated using the Fourier convolution theorem. We assign to
each mesh point a mass from the corresponding density distribution.
The potential is then obtained by a convolution with the Greens
function. We employ the FFT method of \citet{press_etal92} to perform
the convolution.

We pre-compute the individual gravitational potentials $\Phi^{M1}_i$
generated by $J_i$, for unit mass-to-light. This procedure allows us
to quickly obtain the total gravitational potential for any choice of
$M/L_i$ through a weighted sum $\Phi = \sum_i M/L_i \Phi^{M1}_i$,
which is kept constant for each model run.

For a modelling run, we initially tabulate $\Phi_i=M/L_i\Phi^{M1}_i$
on individual Cartesian grids. Forces at grid points are computed by
finite differences. Individual particle accelerations are then
approximated by a cloud-in-cell scheme \citep{hock_east88} to
interpolate the grid point forces to the particle position.

For models in the series M2-M8 we use a spherical harmonic potential
solver as described in \citet{sellwood03}. We use the disc and
spheroid particle populations (particle weights are converted to mass
via the associated $M/L_K$) to calculate the gravitational potential
of the entire particle system. Thus, the contribution of the MD is
neglected.  As discussed at the end of Section~\ref{sec:depro} the
error associated with this approximation is expected to be very small.
For models using the spherical harmonic potential solver, the
potential is updated after every M2M correction step (and temporally
smoothed). We use potential expansion coefficients up to $l_{max}=8$
with $300$ (unevenly spaced) radial bins to $r_{max}=200$~pc. The
width of the innermost bin is $0.2$~pc and of the outermost bin is
$4.2$~pc, which is still smaller than the FWHM of the PSF.

\subsection{Resampling a particle model}
\label{sec:resamp} 

We use the final particle dynamical model as initial conditions for
$N$-body simulations. In order to do this, it is best that the
particles have a narrow range of masses; this ensures both higher
effective mass resolution and a lower artificial two-body relaxation
rate. The models are therefore built using the re-sampling technique
described by \citet{dehnen09}. This section closely follows
\citet{dehnen09} work.  We generate the models using a flat weight
prior $\hat{w}=N^{-1}$. The particle models are re-sampled every $100$
M2M correction steps if the ratio of largest to smallest particle
weight is $\max \{w_i\} / \min \{w_i\} > 10$. Because we do not
normalise total weight (it is only constrained by the observables, in
particular by $A_{00}$), the weight of a re-sampled particle is set to
$w_k = N^{-1}\sum_k w_{i, old}$.  The phase-space coordinates
$(\mathbf{x}_k, \mathbf{v}_k)$ of the k$^{\it th}$ re-sampled orbit
are set to the i$^{\it th}$ original trajectory if
\begin{equation} 
C_i < \bar{\gamma} (k-1/2) \leq C_{i+1}, \;\; i,k \in [1, N] 
\end{equation}
with mean relative normalised weight $\bar{\gamma}=N^{-1}\sum_i
\gamma_i$, cumulative relative normalised weight
$C_i=\sum_{k<i}\gamma_k$ and $\gamma_i=w_i/\hat{w}$ the relative
weight.  Orbits with $\gamma_i<\bar{\gamma}$ are re-sampled at most
once, whereas orbits with $\gamma_i>\bar{\gamma}$ produce at least one
copy. If a trajectory gets re-sampled, the first copy gets the
phase-space position of the original particle. For additional copies
we randomise positions $(x\pm d, y \pm d, z \pm d)$ with $d = (x^2 +
y^2)^{1/2}/100$ exploring the eight distinct combinations of plus and
minus signs. The position of any additional copy is set to $(x + r_x
d, y + r_y d, z + r_z d)$ with $r_{x,y,z} \in [-1,1]$ uniform random
numbers.  We do not alter velocities and every copy keeps the velocity
of the original particle.  At small radii, $r<0.2$ pc, we rotate the
particles randomly about the $z$-axis otherwise the resulting orbits
are too closely spaced.  Our implementation of resampling conserves
total particle number but not individually for the NCD and NCS.  When
a particle is re-sampled more than once all daughter particles inherit
its affiliation to either the NCD or the NCS.

\begin{figure}
\centering
\ifrefereestyle
    \includegraphics[width=0.6\hsize,angle=-90.0]{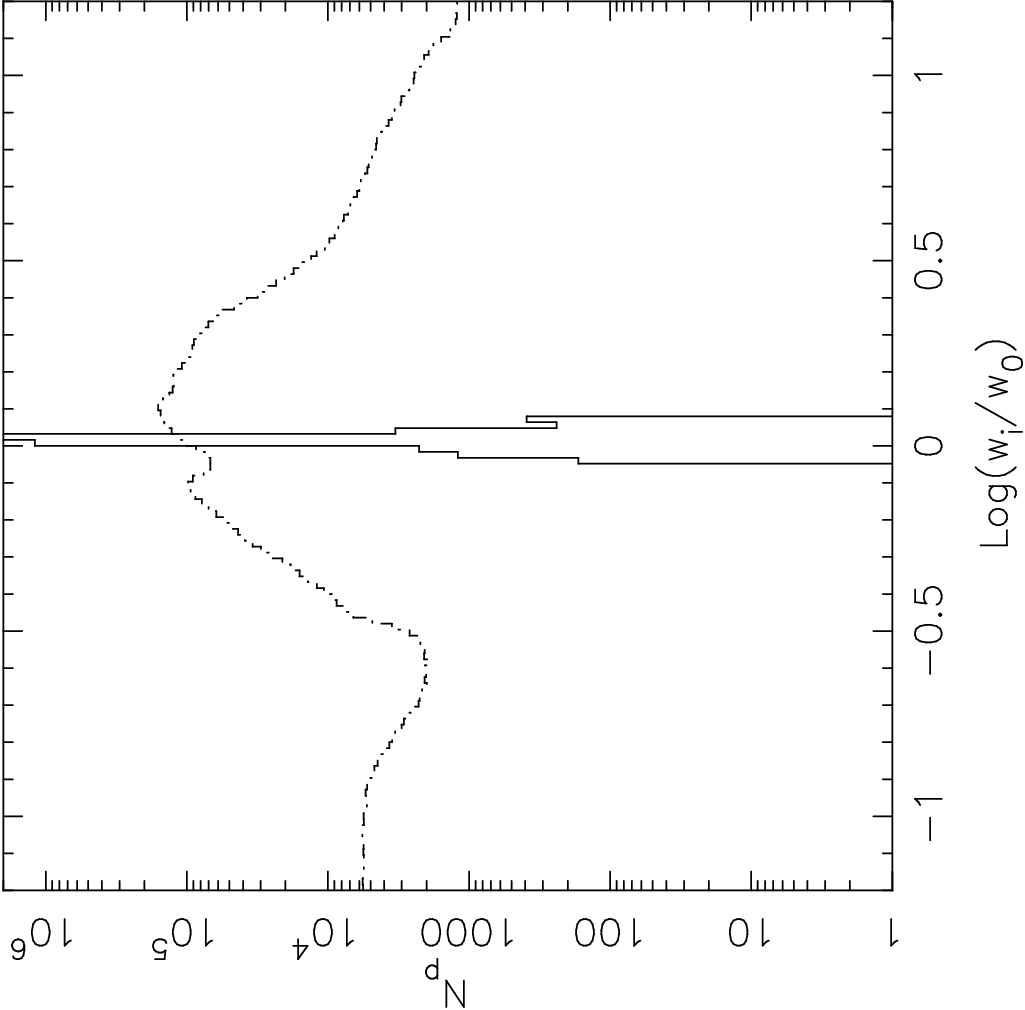}
    \includegraphics[width=0.6\hsize,angle=-90.0]{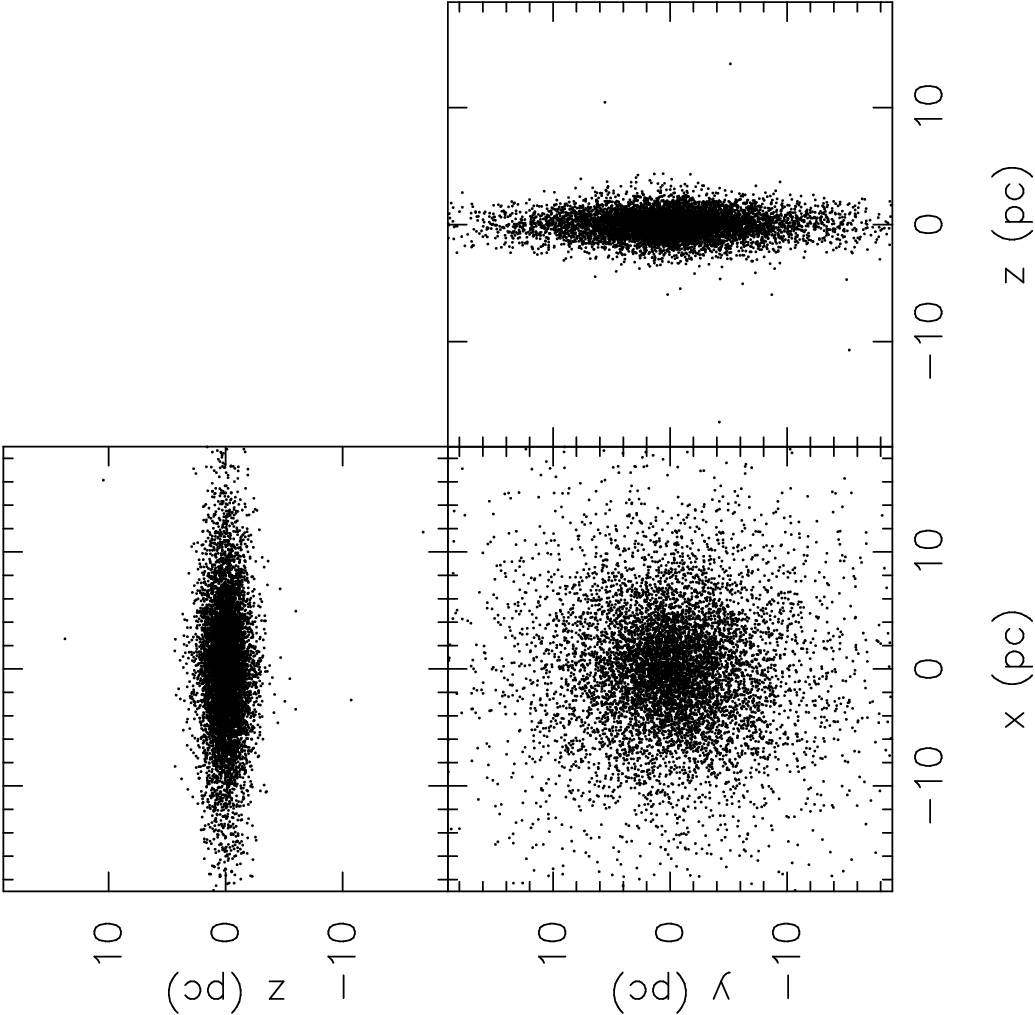}
\else
    \includegraphics[width=\hsize,angle=-90.0]{fig_wdist6M.ps}
    \includegraphics[width=\hsize,angle=-90.0]{fig_coords301split4.ps}
\fi
\vskip0.2truecm
\caption[]{Top: comparison of the particle weight distribution of
  model M2 generated including re-sampling (solid line) and a
  corresponding model produced without re-sampling (dot-dashed line).
  Bottom: Distribution of a subset of NCD particles of the M2 model,
  generated using {\sc nmagic} with re-sampling. The three panels show
  different projections of the particle coordinates. Note that the
  initial NCD model was spherical.}
\label{fig:resamp} 
\end{figure}

Particle re-sampling at work is illustrated in Figure
\ref{fig:resamp}. The top panel compares a final particle weight
distribution of a model generated using {\sc nmagic} without
re-sampling with the peaked distribution of a corresponding model
built including re-sampling. The model with a narrow weight
distribution has a higher effective resolution and hence suffers less
from shot noise than its counterpart with a broad distribution.  The
effective number of particles is defined as $N_{eff}/N =
\overline{w}^2/\overline{w^2}$ \citep{delo+07}. Then, the particle
models shown in Figure \ref{fig:resamp} have $1 - N_{eff}/N = 0.67$
without re-sampling and $1-N_{eff}/N = 9.8 \times 10^{-6}$ with
re-sampling, \ie\ re-sampling leads to $\sim 3\times$ higher
$N_{eff}$.  Starting from a spherical particle population the NCD
shown in the bottom panel is obtained by combining {\sc nmagic} with
re-sampling.

\subsection{Run parameters}

Based on our experience in previous work, we set the value of the
force-of-change parameter $\epsilon=8\times 10^{-8}$ and the temporal
smoothing parameter $\alpha = 2.1\epsilon$.  We set the entropy
parameter $\mu = 2\times 10^{-6}$.  (The parameters
$(\epsilon,\alpha,\mu)$ are defined in \citet{ST96} and
\citet{delo+09}.)  We used timesteps $\delta t = 673$ years, with
$\chi^2$-M2M correction steps every 20 timesteps.  For comparison, a
circular orbit at $r = 13.4$ pc, which contains half the particle
mass, takes 1.82~Myr.


\section{Dynamical models}
\label{sec:models}

In this section we construct dynamical models for the central region
of NGC~4244 to assess its intrinsic kinematics and to constrain the NC
mass. We investigate axisymmetric, two-component models for different
combinations of mass-to-light ratios, fitting the photometry and NIFS
integral field kinematic data.

%
%
All models are constructed including re-sampling of particle
coordinates as described in Section~\ref{sec:resamp}.  In order to
reduce computational cost, the bulk of the modelling is performed
using 0.75M particles, with the gravitational potential calculated on
the Cartesian grids and held fixed throughout. In these models
trajectories are integrated with a standard leapfrog scheme with a
fixed time step.  The models are constructed in a two step process.
First, we start with the spherical isotropic 0.75M particle model and
evolve it using {\sc nmagic} to generate a particle realisation with
desired luminosity distribution, fitting simultaneously but separately
the NCS and NCD photometric constraints. Since only photometric (not
kinematic) constraints are fit at this stage, the velocity scaling is
arbitrary and only the ratio of $M/L_{\rm{NCD}}$ and $M/L_{\rm{NCS}}$
matters for the shape of the gravitational potential. In order to
compute the gravitational potential the ratio
$(M/L_{\rm{NCD}})/(M/L_{\rm{NCS}})$ is fixed at $0.2/1.8$.  The
resulting model then serves as a starting point to simultaneously fit
both the photometric and kinematic constraints ($N_{obs}=2365$
observables) for different combinations of $M/L_{\rm{NCD}}$ and
$M/L_{\rm{NCS}}$.  During this adjustment phase, we typically evolve
for 4M timesteps (2.7 Gyr) and apply 200K M2M correction steps. The
particle system is then relaxed for a further 100K timesteps (67 Myrs)
without changing particle weights. We refer to the final models as
series M1.  We vary $M/L_{\rm{NCS}}$ between $1.0$ and $4.8$, whereas
we use values for $M/L_{\rm{NCD}}$ of $0.1$, $0.2$ and $0.4$. The
influence of the mass-to-light ratio of the MD on the quality of the
fit is expected to be negligible; we therefore keep it constant at
$M/L_{{\rm MD}}=0.7$ compatible with estimates from integrated
colours.

\begin{figure}
\centering 
\ifrefereestyle
    \includegraphics[angle=-90.0,width=0.6\hsize]{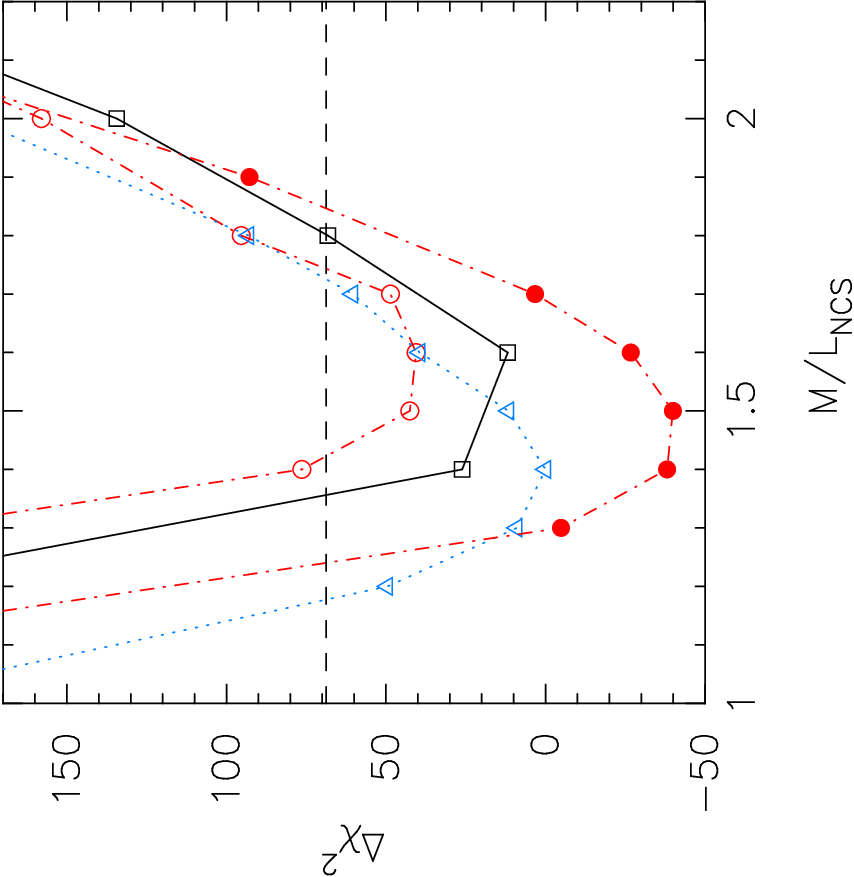}
    \vskip0.2truecm
    \includegraphics[angle=-90.0,width=0.6\hsize]{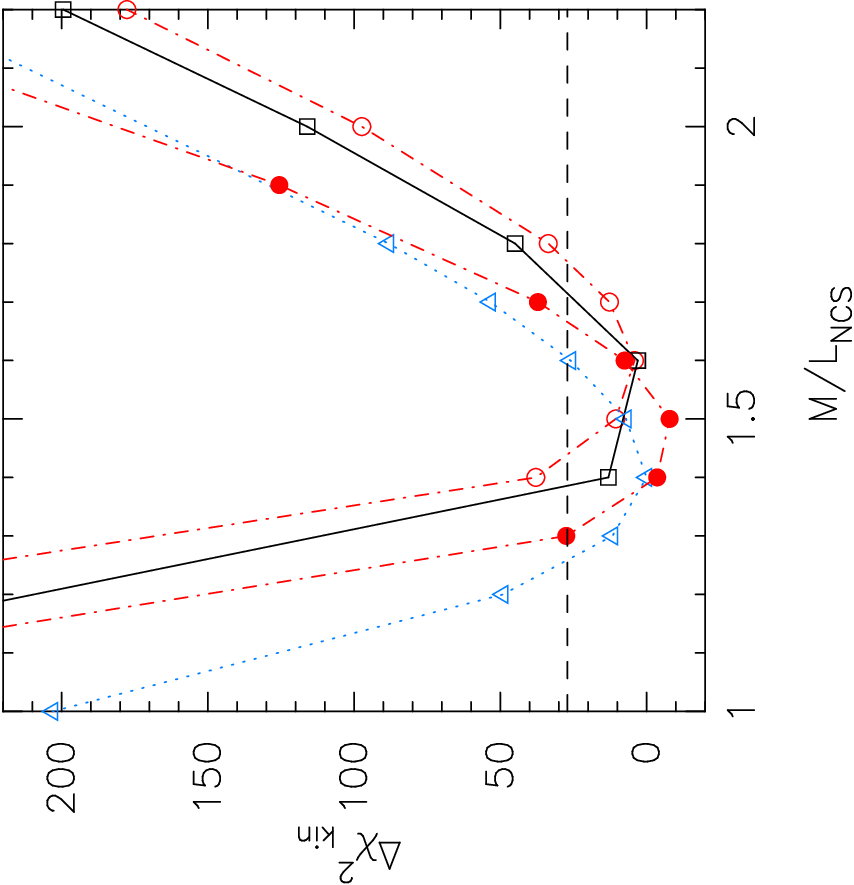}
\else
    \includegraphics[angle=-90.0,width=0.9\hsize]{fig_dchi2tot.ps}
    \vskip0.2truecm
    \includegraphics[angle=-90.0,width=0.9\hsize]{fig_dchi2kin.ps}
\fi
\vskip0.2truecm
\caption[]{$\Delta \chi^2 = \chi^2-\min \{\chi^2\}$ as a function of
  NCS and NCD $K$-band mass-to-light ratio. Upper panel: Total $\Delta
  \chi^2$ for the model fit to the photometric and kinematic target
  observables. Bottom panel: $\Delta \chi^2_{kin}$ of the kinematic
  observables only. Dashed horizontal lines correspond to 68\%
  ($1\sigma$) confidence after \citet{vdBosch_vdVen09}. Symbols are as
  follows: The M1 series for $M/L_{\rm NCD}= 0.1$, $0.2$ and $0.4$ are
  shown as open (red) circles, (black) squares and (blue) triangles,
  joined by dot-dashed, solid and dotted lines, respectively.  Models
  in series M2 are indicated by the solid circles joined by the red
  dot-dashed lines. }
\label{fig:chi2}
\end{figure}

The results are presented in Figure~\ref{fig:chi2}, which illustrates
how the quality of the model fit changes with mass-to-light ratios.
The top panel shows $\Delta\chi^2=\chi^2-\min\{\chi^2\}$, whereas
$\Delta\chi_{kin}^2$ of the kinematic observables alone
($\Delta\chi^2$ "marginalised" over the $A_{lm}$'s) is shown in the
bottom panel. The 68\% confidence limits ($1\sigma$) are computed as
$\sqrt{2N_{obs}}$ following \citet{vdBosch_vdVen09}. These limits are
$\Delta\chi^2=68.8$ and $\Delta\chi_{kin}^2=27.0$, respectively.

Regardless of whether $\Delta\chi_{kin}^2$ or $\Delta\chi^2$ is used,
the resulting best model has mass-to-light ratios $M/L_{\rm NCD}=0.4$
and $M/L_{\rm NCS}=1.4$. Note that the models with $M/L_{\rm NCD}=0.1$
or $M/L_{\rm NCD}=0.2$ and $M/L_{\rm NCS}=1.6$ reproduce the NIFS data
with comparable quality and are in better agreement with $M/L_{\rm
  NCD}=0.1-0.25$ estimated from integrated colours (see
Section~\ref{sec:photo}) than is $M/L_{\rm NCD}=0.4$.  Moreover,
$M/L_{\rm NCD}=0.1$ also agrees with the estimates from Bruzual \&
Charlot models \citep{bruzual_charlot03}.

The range of acceptable NCS masses is estimated as the range for which
$\Delta\chi^2$ ($\Delta\chi_{kin}^2$) is below $1 \sigma$ confidence
after "marginalising" over $M/L_{\rm NCD}$. Although $\Delta\chi^2$ is
the more appropriate quantity to discriminate between models because
both photometric and kinematic constraints are imposed on the models,
the allowed NCS mass ranges determined using either $\Delta\chi^2$ or
$\Delta\chi_{kin}^2$ agree with each other. We obtain a NCS mass of
$M_{\rm NCS}=1.6^{+0.5}_{-0.2}\times 10^7~ \Msun$ within $\approx
42.4$~pc.  This is almost an order of magnitude higher than the lower
limit of $\sim$2.5$\times$10$^6~ \Msun$ obtained from the observed
velocity of an H{\sc ii} region at a projected distance of 19 pc from
the NC center \citep{seth06}. The mass within $\sim 15$ pc is $1.0
\times 10^7~\Msun$, which agrees with the mass within the same radius
obtained from the {\sc jam} models in H11, $(1.1 \pm 0.2)\times
10^7~\Msun$.  The mass of the NCD is not as well constrained: we
obtain $3.6 \times 10^5~ \Msun$ for $M/L_{\rm NCD}=0.1$ and $14.4
\times10^5~ \Msun$ for $M/L_{\rm NCD}=0.4$.

%
%

At this point models with 6M particles are constructed starting from the
spherical isotropic 6M particle initial conditions. We start by generating high
resolution versions of models M1 with $M/L_{\rm NCD}=0.1$. We again use
photometric followed by photometric$+$kinematic constraints. But now we replace
the {\sc FFT} method with a spherical harmonics potential solver in order to
obtain higher spatial resolution at the centre. These models also use a
Runge-Kutta time-integrator with adaptive timestep (using routine {\sc odeint}
of \citet{press_etal92} with accuracy parameter $10^{-6}$) to allow a
comparison with the models which include a black hole, presented in the next
section below. Since the potential is computed via an expansion in $l, m$
spherical harmonics analogous to the photometric constraints, we include the
same terms in the expansion of the potential as for the luminosity density (\ie
non-zero $A^{\rm NCD}_{lm}$ and $A^{\rm NCS}_{lm}$, \cf also
Section~\ref{ssec:pot}).  The potential is recalculated after every M2M
correction step. We refer to the resulting models as series M2.

The 6M particle models illustrate several interesting points. The NCS
mass estimated using 6M particles agrees with the estimates presented
above using the M1 models. This suggests that the inferred NCS mass is
robust with respect to how the models have been constructed, in
particular to the potential solver, integration scheme and number of
particles. Increasing the number of particles from M1 to M2 decreases
$\chi^2_{kin}$ of the corresponding best models by only a small amount
(if at all) with respect to the $\Delta\chi^2_{kin}$ confidence limit.
This indicates that the model fit to the NIFS data is dominated by the
uncertainties in the data while the contribution of shot noise to
$\chi^2_{kin}$ is negligible. On the other hand, $\chi^2$ is reduced
considerably mainly due to a reduction in $\chi^2_{Alm}$\footnote{We
  use the same MC $A_{lm}$ errors as for models M1.}. This suggests
that $\chi^2_{Alm}$ is dominated by Poisson noise.  Generally
$\chi^2_{Alm} < N_{Alm}$ because of the temporal smoothing, and both
the M1 and M2 models reproduce the photometric data very well.

\begin{figure}
\centering
\ifrefereestyle
    \includegraphics[width=0.8\hsize,angle=0.0]{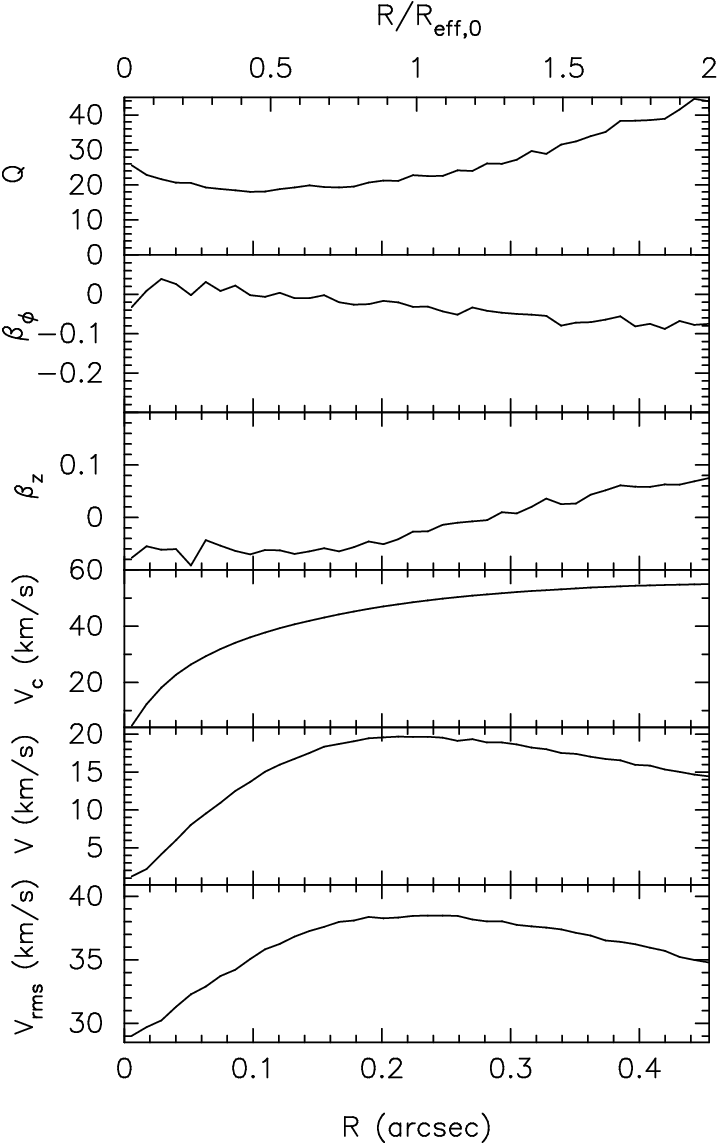}
\else
    \includegraphics[width=0.9\hsize,angle=0.0]{kinematics_initial.ps}
\fi
\vskip0.2truecm
\caption[]{The intrinsic kinematics of the NC in the best-fit model in
  the M2 series.  From bottom up we show the second moment of
  line-of-sight velocity $V_{rms}$, line-of-sight velocity $V$, the
  circular velocity $V_c$, the vertical anisotropy $\beta_z$, and the
  tangential anisotropy $\beta_\phi$.  The top row shows the
  Toomre-$Q$ of the NCD only.}
\label{fig:intkin}
\end{figure}

We present the intrinsic kinematics of the best-fit model in series M2
in Figure \ref{fig:intkin}.  We computed the radial, tangential and
vertical dispersions, $\sigma_u$, $\sigma_v$ and $\sigma_w$
respectively, and plot the anisotropies $\beta_\phi = 1 -
(\sigma_v/\sigma_u)^2$ and $\beta_z = 1 - (\sigma_w/\sigma_u)^2$.  In
agreement with H11 we find that $\beta_z < 0$.  Also $V_{rms}$ has a
central minimum.  H11 found that the combination of these two
properties provide important constraints on the amount of mass that
the nuclear cluster could have accreted in the form of star clusters,
as we shall also see below.

\subsection{Adding intermediate mass black holes}

Some galaxies are known to harbour both a NC and a massive black hole
\citep{seth08,graham+spitler_09b,neumayer_etal_12}.  Since mass and anisotropy
are degenerate with each other \citep{binney_mamon_1982}, we wish to explore
how adding an IMBH might change $\beta_z$.  We therefore also generate models
including an IMBH at their centre with the aim of finding a robust upper IMBH
mass limit compatible with the observations.  We construct these models in a
manner analogous to models M2 above, using a spherical harmonics potential
solver in order to obtain higher spatial resolution at the centre. These models
again use a Runge-Kutta time-integrator with adaptive timestep for higher
accuracy in the vicinity of the IMBH.  Using two-integral {\sc jam} models
\citep{cappellari08} H11 obtained an upper limit of $\Mbh \la 10^5~\Msun$ on
any black hole that may be present.  We revisit this estimate with our more
general three-integral modelling.

\begin{figure*}
\begin{tabular}{ccc}
\includegraphics[angle=-90.0,width=0.3\hsize]{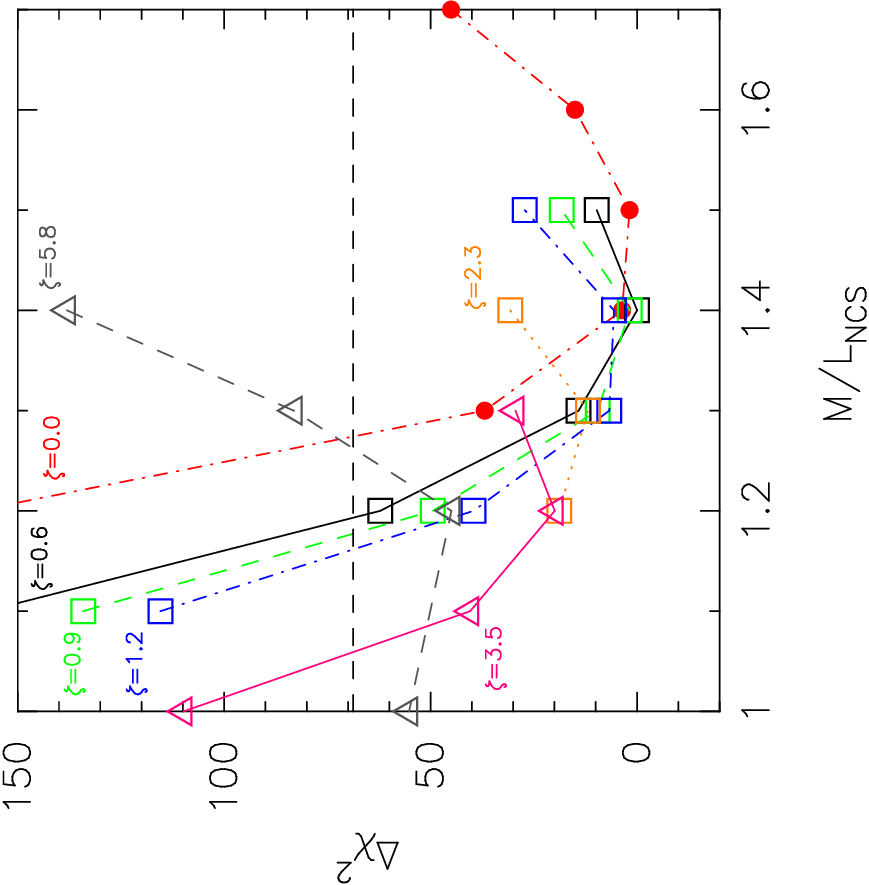} &
\includegraphics[angle=-90.0,width=0.3\hsize]{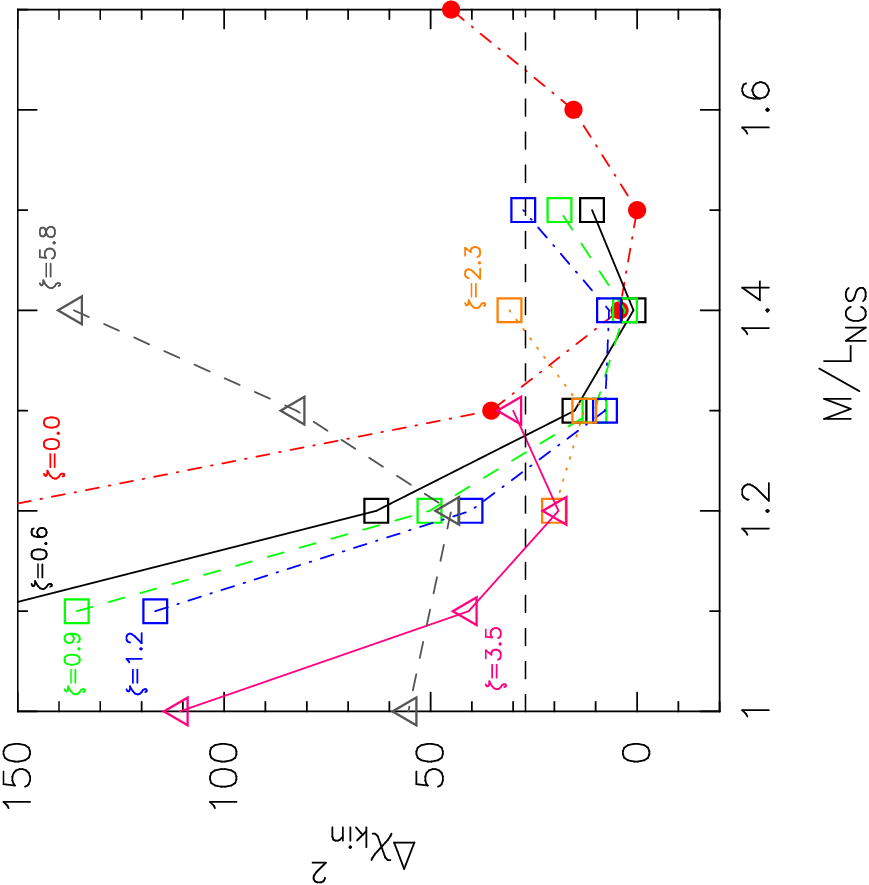} &
\includegraphics[angle=-90.0,width=0.3\hsize]{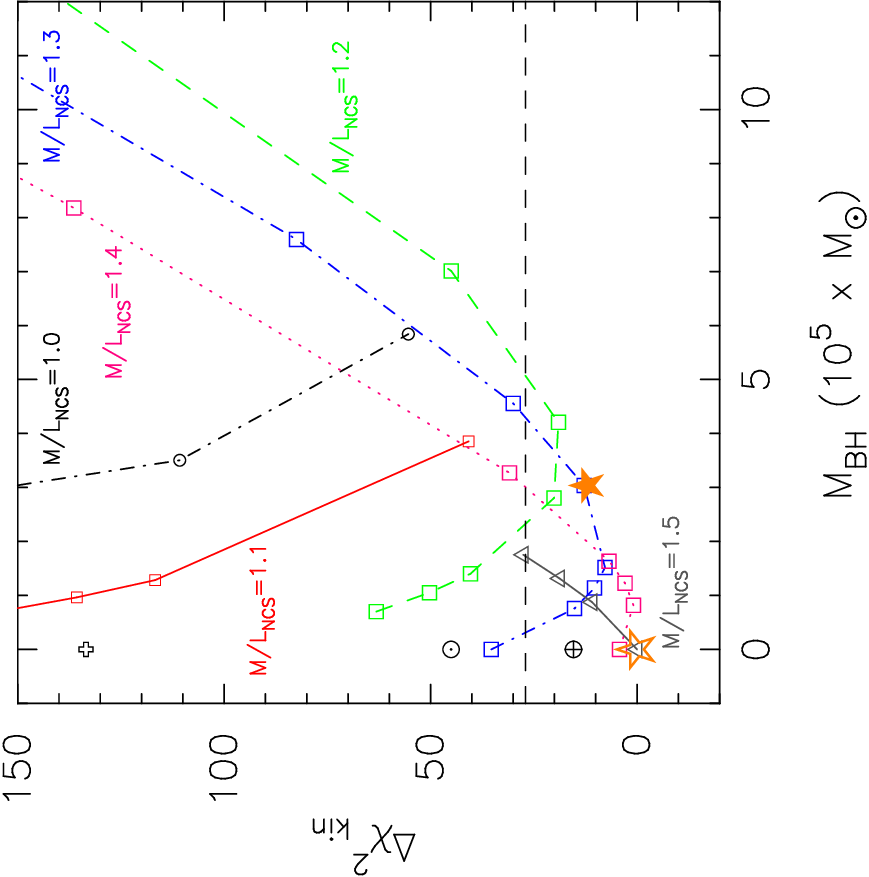} \\
\end{tabular}
\caption[]{%
  $\Delta \chi^2 = \chi^2-\min \{\chi^2\}$ as a function of NCS
  mass-to-light ratio for various IMBH mass fractions $\zeta = \Mbh /
  (M/L_{NCS})$.  Left panel: Total $\Delta \chi^2$ for the model fit
  to the photometric and kinematic target observables. Middle panel:
  $\Delta \chi^2_{kin}$ of the kinematic observables only. Right
  panel: $\Delta \chi^2_{kin}$ as a function of \Mbh\ instead of
  $M/L_{\rm NCS}$.  Dashed horizontal lines correspond to 68\%
  ($1\sigma$) confidence after \citet{vdBosch_vdVen09}. In the left
  and middle panels the symbols are as follows: The M2 series is shown
  as solid (red) circles joined by red dot-dashed lines. Models
  including a central IMBH are shown as open squares joined by (black)
  solid (M3), (green) dashed (M4), (blue) dot-dashed (M5) or (orange)
  dotted (M6) lines. The open triangles are for series M7 and M8.  The
  M7 and M8 series are joined by (pink) solid and (grey) dashed lines
  respectively. IMBH mass fractions for models in the M3 to M8 series
  are given in Table \ref{tab:runs}. In the right panel the different
  symbols represent: circles joined by (black) dot-dashed line:
  $M/L_{\rm NCS} = 1$, squares joined by (red) solid line: $M/L_{\rm
    NCS} = 1.1$, squares joined by (green) dashed line: $M/L_{\rm NCS}
  = 1.2$, squares joined by (blue) dot-dashed line: $M/L_{\rm NCS} =
  1.3$, squares joined by (pink) dotted line: $M/L_{\rm NCS} = 1.4$,
  triangles joined by (grey) solid line: $M/L_{\rm NCS} = 1.5$,
  (single) crossed circle: $M/L_{\rm NCS} = 1.6$, (single) dotted
  circle: $M/L_{\rm NCS} = 1.7$, (single) cross: $M/L_{\rm NCS} =
  1.9$, and (single) diamond: $M/L_{\rm NCS} = 2.2$. The open and
  solid (orange) stars in the middle and right panels mark the best
  fit M2 and M6 models, respectively.}
\label{fig:chi2_BH}
\end{figure*}

The IMBH is represented by a Plummer potential with scale-length set
to $0.02$~pc. We generate models M3 to M8 for various IMBH mass
fractions.  The IMBH mass fractions $\zeta = \Mbh /(M/L_{NCS} \times
10^5 \Msun)$ are given in Table~\ref{tab:runs}. For each series of
models the black hole mass is then given as $\Mbh = \zeta \times
M/L_{\rm NCS}$.  The results of the models are presented in
Figure~\ref{fig:chi2_BH}, which illustrates how the quality of the
model fit changes with $M/L_{\rm NCS}$ and $\zeta$. The left panel
shows $\Delta\chi^2 = \chi^2-\min\{\chi^2\}$, whereas
$\Delta\chi_{kin}^2$ of the kinematic observables alone is shown in
the middle and right panels. The 68\% confidence limits are as in the
above \ie\ $\Delta\chi^2=68.8$ and $\Delta\chi_{kin}^2=27.0$.

As expected, the minimum $\chi^2$ along a given line in
Figure~\ref{fig:chi2_BH} shifts towards smaller $M/L_{\rm NCS}$ with
increasing IMBH mass fraction (see especially the right panel of
Figure \ref{fig:chi2_BH}). Each line in Figure \ref{fig:chi2_BH}
intersects the confidence limit (dashed horizontal line) twice (in the
case of series M3 and M4 the modelling sets need to be extrapolated).
The intersection with larger $M/L_{\rm NCS}$ corresponds to the
largest admissible IMBH mass along each line. The largest IMBH mass
compatible with the data would be obtained at a line that intersects
the horizontal line only once, at its minimum $\chi^2$.  Using
$\Delta\chi^2_{kin}$ shown in the right panel of
Figure~\ref{fig:chi2_BH}, the dashed line with $M/L_{\rm NCS}=1.2$
leads to the IMBH mass upper limit of $\Mbh = 5.0 \times 10^5\,\Msun$.
This upper limit is larger than the one found in H11 using JAM models,
presumably reflecting the greater orbital freedom presented by
3-integral versus 2-integral models.

\begin{centering}
\begin{table*}
\vbox{\hfil
\begin{tabular}{ccccccc}
  \hline
  \\[-1.7ex]
  \textsc{Series} & $N_p$ ($\times 10^{6}$) & $M/L_{\rm NCD}$ & 
  $M/L_{\rm NCS}$ & $\zeta = \frac{M_{\rm BH}}{(M/L_{\rm NCS}) \times 10^5 M_{\odot}}$
& \textsc{Pot.} & \textsc{Int.} \\
  \\[-1.7ex] 
  \hline
  \textsc{M1} & $0.75$ & $0.1, 0.2, 0.4$ & $1.0-4.8$ & $-$ &
\textsc{FFT} & \textsc{Leapfrog}  \\
  \textsc{M2} & $6$ & $0.1$ & $1.0-2.2$ & $-$ &
\textsc{Sph. harm.} & \textsc{Runge-Kutta} \\
  \textsc{M3} & $6$ & $0.1$ & $1.1-1.5$ & $0.6$ &
\textsc{Sph. harm.} & \textsc{Runge-Kutta} \\
  \textsc{M4} & $6$ & $0.1$ & $1.1-1.5$ & $0.9$ &
\textsc{Sph. harm.} & \textsc{Runge-Kutta} \\
  \textsc{M5} & $6$ & $0.1$ & $1.1-1.5$ & $1.2$ &
\textsc{Sph. harm.} & \textsc{Runge-Kutta} \\
  \textsc{M6} & $6$ & $0.1$ & $1.2-1.4$ & $2.3$ &
\textsc{Sph. harm.} & \textsc{Runge-Kutta} \\
  \textsc{M7} & $6$ & $0.1$ & $1.0-1.3$ & $3.5$ &
\textsc{Sph. harm.} & \textsc{Runge-Kutta} \\
  \textsc{M8} & $6$ & $0.1$ & $1.0-1.4$ & $5.8$ &
\textsc{Sph. harm.} & \textsc{Runge-Kutta} \\
  \hline
\end{tabular}
\hfil}
\caption{%
  Summary of the modelling runs. Columns from left to right are:
  Name of the series of models, number of particles, mass-to-light 
  ratio of the NCD, mass-to-light ratio of the NCS, IMBH mass fraction 
  $\zeta$, the method used to compute the potential and the time 
  integration scheme.}
\label{tab:runs}
\end{table*}
\end{centering}

If we use $\Delta\chi^2$ shown in the left panel of
Figure~\ref{fig:chi2_BH} instead of $\Delta\chi^2_{kin}$, we find that
even larger IMBH masses (up to a factor $2$ or higher) are compatible
with the data.  Nonetheless, we use the more conservative IMBH mass
range provided by the NIFS data alone.

\begin{figure*}
\centering
\includegraphics[width=0.75\hsize,angle=-90.0]{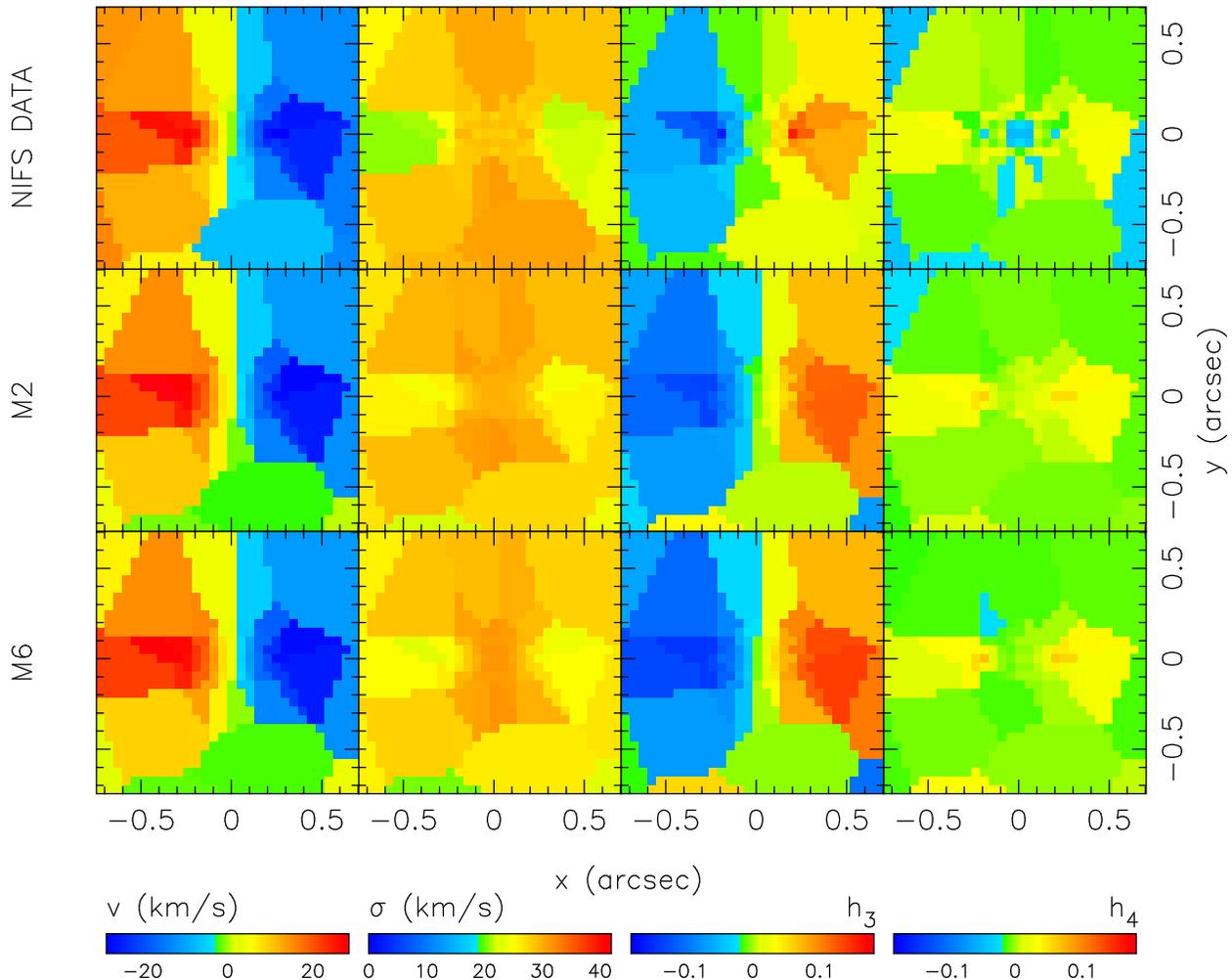}
\vskip0.2truecm
\caption[]{Symmetrised NIFS integral-field kinematic data within $\pm
  0.7 \arcsec$ of the central region of NGC~4244 (top row) compared
  with corresponding luminosity-weighted data extracted from the
  best-fit models in series M2 (middle row) and M6 with $M/L_{\rm
    NCD}=0.1$, $M/L_{\rm NCS}=1.3$ and $Mbh = 3.0 \times
  10^5\,\Msun$ (bottom row).  These models are indicated by the open
  and solid orange stars, respectively, in the right panel of
  Figure~\ref{fig:chi2_BH}.  From left to right are shown:
  line-of-sight velocity $v$, line-of-sight velocity dispersion
  $\sigma$ and higher order Gauss-Hermite moments $h_3$ and $h_4$.}
\label{fig:nifskin_model} 
\end{figure*}

\begin{figure}
\centering
\includegraphics[width=0.9\hsize,angle=-90.0]{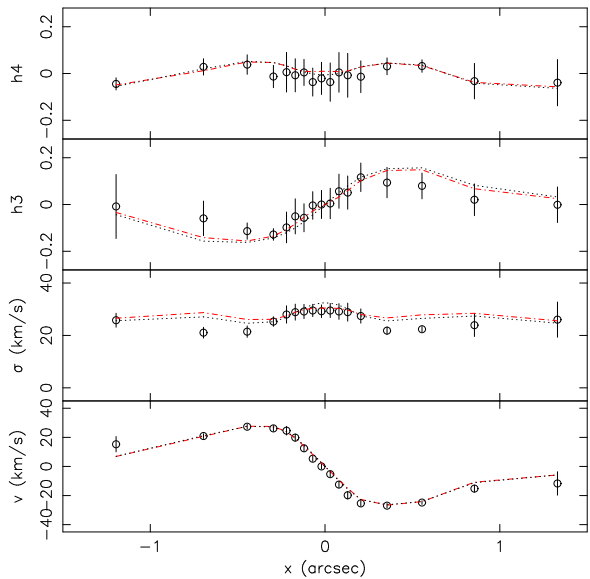}
\vskip0.2truecm
\caption[]{Comparison of the best-fit models in series M2 and M6
  (indicated by the open and solid orange stars in
  Figure~\ref{fig:chi2_BH}) to the kinematic data along the major-axis
  extracted from the NIFS data. Models and data are shown as lines and
  open circles, respectively, with M2 indicated by the dot-dashed
  (red) line and M6 by the dotted (black) line. From bottom to top are
  shown: velocity $v$, velocity dispersion $\sigma$, and Gauss-Hermite
  moments $h_3$ and $h_4$. The model kinematic data were computed using
  a Gauss-Hermite fit to the velocity distribution along the
  line-of-sight.}
\label{fig:mj_kin} 
\end{figure}

Figure~\ref{fig:nifskin_model} shows a comparison of the best fit
models in series M2 and M6 (indicated by the orange stars in the
middle and right panels of Figure~\ref{fig:chi2_BH}) with the integral
field NIFS kinematic. The model fits to the NIFS data are excellent.
$\chi^2_{kin}$ for the best models M2 and M6 are $194.715$ and
$207.532$, respectively. Figure~\ref{fig:mj_kin} shows a comparison of
the best fit models in the M2 and M6 series with major-axis kinematic
data extracted from the NIFS data.  The model kinematics are computed
from a Gauss-Hermite fit to the line-of-sight velocity distribution in
the corresponding Voronoi bins.  To compute the temporally smoothed
LOSVD, we use $27$ bins in velocity within a range of width $300$
\kms, centred on the corresponding NIFS line-of-sight velocity. Note
that we did not fit the full LOSVD itself, instead we constrained the
particle models using luminosity-weighted moments as described in
Section~\ref{sec:obs}.

\section{$N$-body Simulations}
\label{sec:nbody}

Our best dynamical model without an IMBH is M2 with $M/L_{\rm
  NCS}=1.5$.  After building this model we used it for a number of
$N$-body experiments testing its stability and evolution by using it
as initial conditions.  The simulations, were evolved with {\sc
  pkdgrav} \citep{stadel_phd}, an efficient, multi-stepping, parallel
treecode.  In all cases we use an opening angle $\theta = 0.7$.  We
used base timestep $\Delta t = 0.1$ Myr and changed timesteps of
individual particles such that $\delta t = \Delta t/2^n < \eta
(\epsilon/a)^{1/2}$, where $\epsilon$ is the softening and $a$ is the
acceleration of the particle, with n as large as 29 allowed.  We set
$\eta = 0.03$, a quite conservative value.

\subsection{Stability test}

In constructing our dynamical model we have assumed that the NC in NGC
4244 is very likely axisymmetric.  H11 found no evidence of
non-axisymmetry in the NC of M33. Its PA is consistent with that of
its main disc and its apparent ellipticity is consistent with a
vertical flattening of $q=0.7$, the average observed in the NCs of
edge-on late-type galaxies \citep{seth06}.  There is also only a small
misalignment between the photometric and the kinematic
major axes. At present, M33 is the only galaxy in which the axial
symmetry of the NC can be determined.

Our first $N$-body simulation therefore tests the stability of the model
against non-axisymmetric perturbations, particularly the bar instability, which
plagues rapidly rotating systems.  After evolving the best fit model in series
M2 for 50 Myr (the rotation period at 5 pc being 0.63 Myr), the model remained
axisymmetric with no hint of a bar or spirals.  The top panel of Figure
\ref{fig:intkin} plots the Toomre-$Q = \sigma_u\kappa/(3.36G\Sigma)$ of the
NCD, where $\kappa$ is the epicyclic radial frequency, $G$ is the gravitional
constant and $\Sigma$ is the surface density.  The stability of the system
stems from the high Toomre-$Q$ of the NCD, which is everywhere greater than
$10$.

\subsection{Accretion simulations}
\label{ssec:accretionsims}

H11 explored the hypothesis that NC formation is a result of star
cluster (SC) accretion.  They showed that the observed kinematics of
the NC in NGC 4244 are not consistent with accretion of more than
$\sim 50\%$ of its mass in the form of SCs.  Specifically, once the
accreted mass fraction exceeded this value the resulting central
$V_{rms}$ was no longer a minimum.  We will see here too that
$V_{rms}$ starts to lose its central minimum once the accreted mass
fraction becomes too large, and our constraint is even more stringent
than that of H11.  On the other hand, H11 showed that the negative
$\beta_z$ of the NC was only possible if it accreted $\ga 10\%$ of its
mass as SCs on highly inclined orbits.  Those simulations assumed SCs
accreting onto a pre-existing NC, either as a NCD or as an isotropic
NCS.  Therefore we next subject our best-fitting model of the NC to SC
accretions.

In order to permit the model star clusters to sink to the centre via
dynamical friction, we introduce the best fit model in the M2 series
inside a particle main disc with an exponential profile, because
NGC~4244 is a late-type, bulgeless galaxy. We use the same model for
the MD as did H11, \ie\ four million multi-mass particles with masses
ranging from $7\Msun$ within the inner $20~\pc$ increasing to
$1.2\times10^7~\Msun$ in the disc's outskirts.  The distributions of
masses and softenings of the main disc particles are shown in Figure 6
of H11; the softening is related to particle mass via $\epsilon_p
\propto m_p^{1/3}$.

We accrete three of the model SCs described in H11 which we term G1,
G2 and G3 in order of increasing mass.  Their properties are listed in
Table \ref{tab:scmodels}.  As in H11, we define the concentration of
each star cluster as $c\equiv \log(\re/R_c)$ where \re\ is the half
mass radius (effective radius) and $R_c$ is the core radius, where the
surface density drops to half of the central.  These were allowed to
accrete onto the NC starting from circular orbits at $50~\pc$ from the
centre.  We start the SCs from 4 different inclinations relative to
the NCD: $0\degrees$, $30\degrees$, $60\degrees$, and $90\degrees$.
The SCs require about 40 Myrs to accrete onto the NC.

\begin{figure*}
\centering
\begin{tabular}{ccc}
\includegraphics[width=0.3\textwidth]{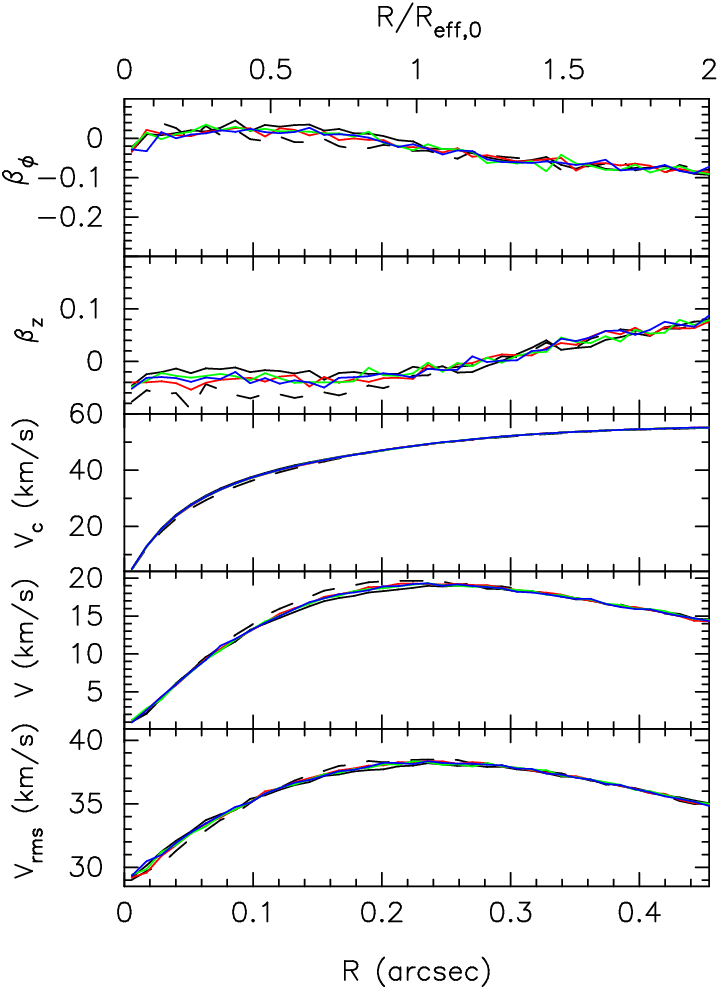} &
\includegraphics[width=0.3\textwidth]{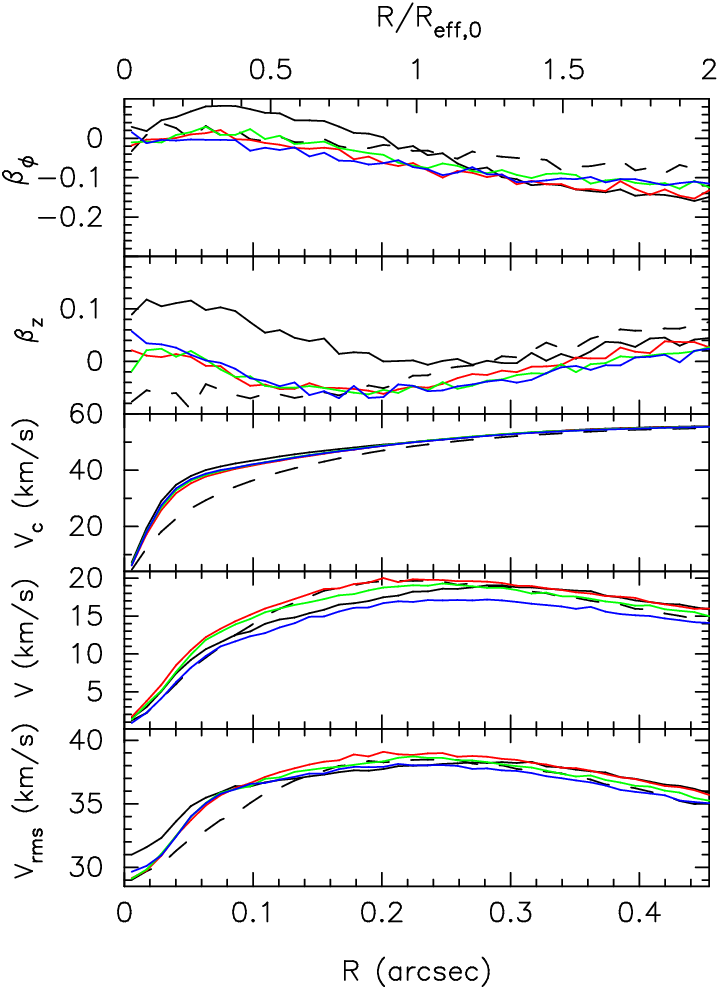} &
\includegraphics[width=0.3\textwidth]{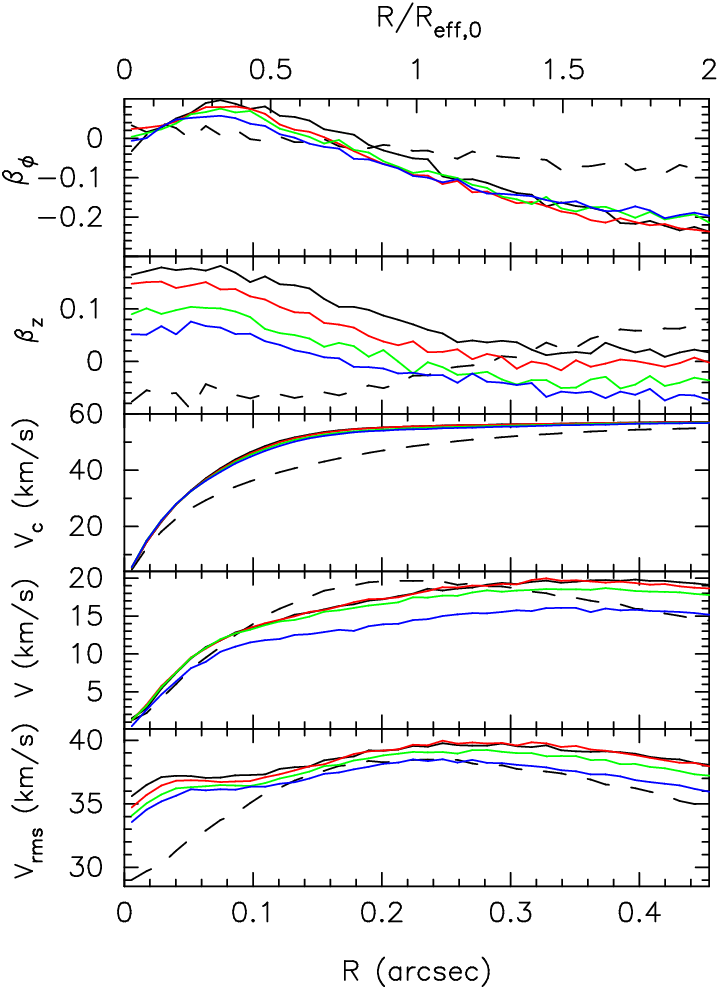} \\
\end{tabular}
\caption{The effect of accreting star clusters onto the NC in NGC
  4244.  From left to right these show the effect of accreting SCs G1,
  G2 and G3.  The black, red, green and blue solid lines show SCs
  accreting from $0\degrees$, $30\degrees$, $60\degrees$ and
  $90\degrees$, respectively.  The dashed line shows the initial NC.}
\label{fig:accretions}
\end{figure*}

The results of these accretion simulations are shown in Figure
\ref{fig:accretions}.  The low density star cluster G1 is disrupted at
$\sim 8$ pc from the centre.  Thus it barely perturbs the kinematics
of the NC.  However we note that there is a general tendency for
$\beta_z$ to increase slightly within \re, suggesting that even this
mild $\sim 1\%$ mass accretion can alter the kinematics.  The more
massive, denser, clusters G2 and G3, both of which sink all the way to
the centre, perturb the kinematics much more.  In all cases $\beta_z$
increases; for G2 $\beta_z$ averaged within \re\ is nearly zero, while
$\beta_z > 0$ everywhere within \re\ when G3 is accreted.

Accreting G3 (which has a mass $\sim 13\%$ that of the NC) also raises
the central \Vrms.  Although \Vrms\ at the centre remains smaller than
at \re, there does not seem to be much room for further significant
accretions without making \Vrms\ centrally peaked, unlike the
observations. This is likely to hold also if this mass fraction
arrives as many smaller star clusters, provided that the star clusters
are dense enough that some fraction of their stars survives all the
way to the centre of the NC.

\begin{table}
\begin{centering}
\begin{tabular}{ccccc} \\ \hline
\multicolumn{1}{c}{Model} &
\multicolumn{1}{c}{$\rm M_*$} &
\multicolumn{1}{c}{\re} &
\multicolumn{1}{c}{$c$} &
\multicolumn{1}{c}{H11 name} \\
  & [$\times 10^5$\Msun]   & [pc] & \\ \hline
G1   & $2$ & 1.11 & 0.12 & C4 \\
G2   & $6$ & 1.11 & 0.16 & C5 \\
G3   & $20$ & 2.18 & 0.12 & C3 \\
\hline
\end{tabular}
\caption{The star clusters used in the accretion simulations.  $\rm
  M_*$ is the stellar mass of the SC, \re\ is the effective
  (half-mass) radius, and $c$ is the concentration (defined in the
  text).  For comparison, the last column lists the name used for the
  model in H11.}
\label{tab:scmodels}
\end{centering}
\end{table}

\section{Discussion and conclusions}
\label{sec:conclusions}

\begin{figure}
\centering
\vskip0.4truecm
\includegraphics[width=0.8\hsize,angle=-90.0]{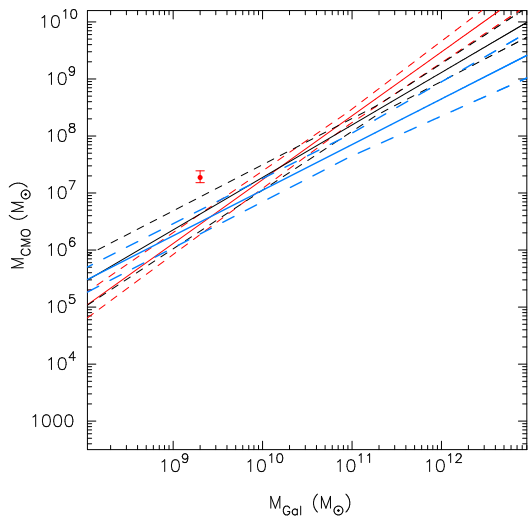}
\caption[]{Mass of the NC of NGC 4244 versus the mass of the host
  galaxy compared with the $M_{CMO}-M_{gal}$ relations of
  \citet{fer_etal_06}. The solid red and black lines show the
  correlations for NCs and SMBHs in early-type galaxies, respectively,
  with 1 $\sigma$ confidence levels shown as dashed lines.  The blue
  lines show the relation of \citet{scott_graham12}.}
\label{fig:ferrarese}
\end{figure}

We have performed a dynamical study of the nuclear cluster in the
edge-on spiral galaxy NGC~4244 taking into account different
morphological components, which are the galaxy main disc (MD), nuclear
cluster disc (NCD) and nuclear cluster spheroid (NCS).  We have
constructed axisymmetric dynamical particle models accounting for the
MD, the NCD and the NCS. We find a total NCS mass $M_{\rm NCS} =
1.6^{+0.5}_{-0.2}\times 10^7~ \Msun$ within approximately $42.4$~pc
($2\arcsec$).  Both the fits of \citet{seth05a} and of
\citet{fry+1999} show that there is no obvious bulge component in NGC
4244.  Using the {\it 2MASS} Large Galaxy Atlas \citep{jarrett+2003}
$K$-band magnitude, the total luminosity of the galaxy is $3.2 \times
10^9~ \Lsun$, and thus the galaxy stellar mass is $~ 2 \times 10^9 ~
\Msun$.  Figure \ref{fig:ferrarese} plots the NC mass compared with
the $M_{CMO}-M_{gal}$ relation.  The NC sits above this relation.

\subsection{Vertical anisotropy}

The kinematics are moderately tangentially anisotropic inside \re\
with an anisotropy parameter $\beta_z \sim -0.1$.  This is in good
agreement with the {\sc jam} models of H11.  H11 showed that $\beta_z
< 0$ requires high-inclination infall of star clusters onto a
pre-existing nuclear cluster.  In our accretion simulations onto a
more realistic model of the NC we found that even the accretion of a
star cluster of just $13\%$ the mass is enough to erase the vertical
anisotropy.  This raises questions about whether such anisotropy can
be due to accretion at all.  It also hints that, unless we are
observing the NC of NGC 4244 at a special time, it cannot sustain
accretion of $\ga 10\%$ mass as suggested by H11.

\begin{figure}
\centering
\includegraphics[width=0.7\hsize,angle=-90.0]{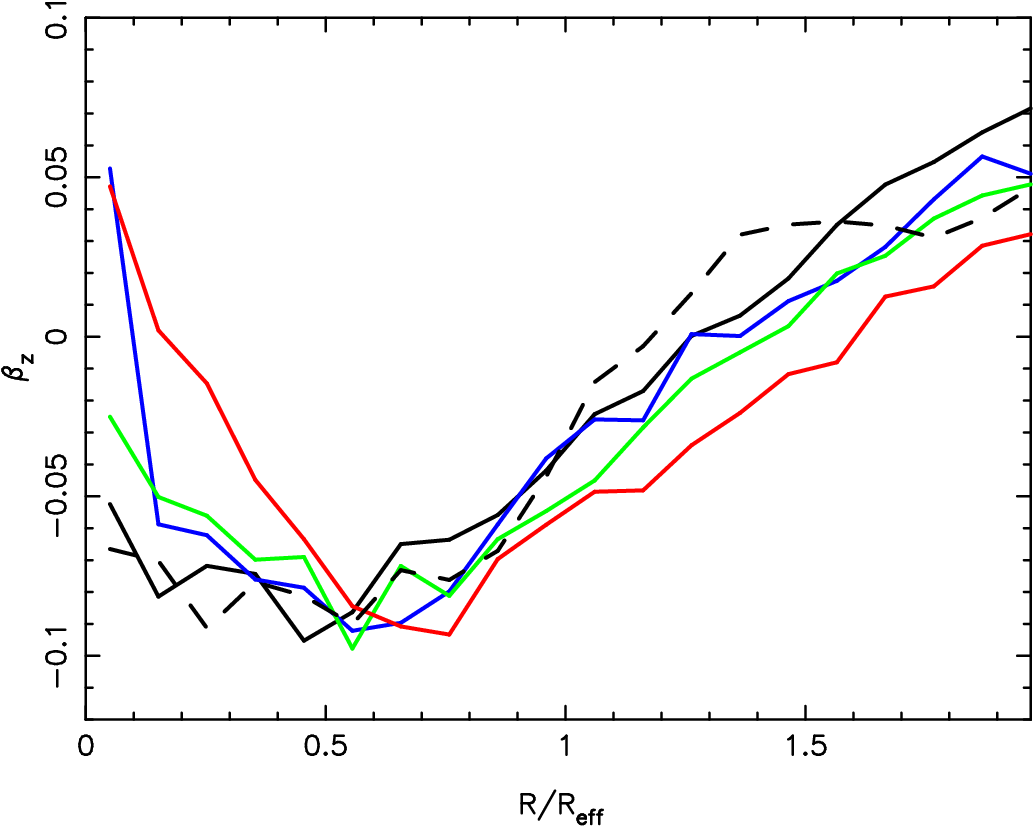}
\includegraphics[width=0.7\hsize,angle=-90.0]{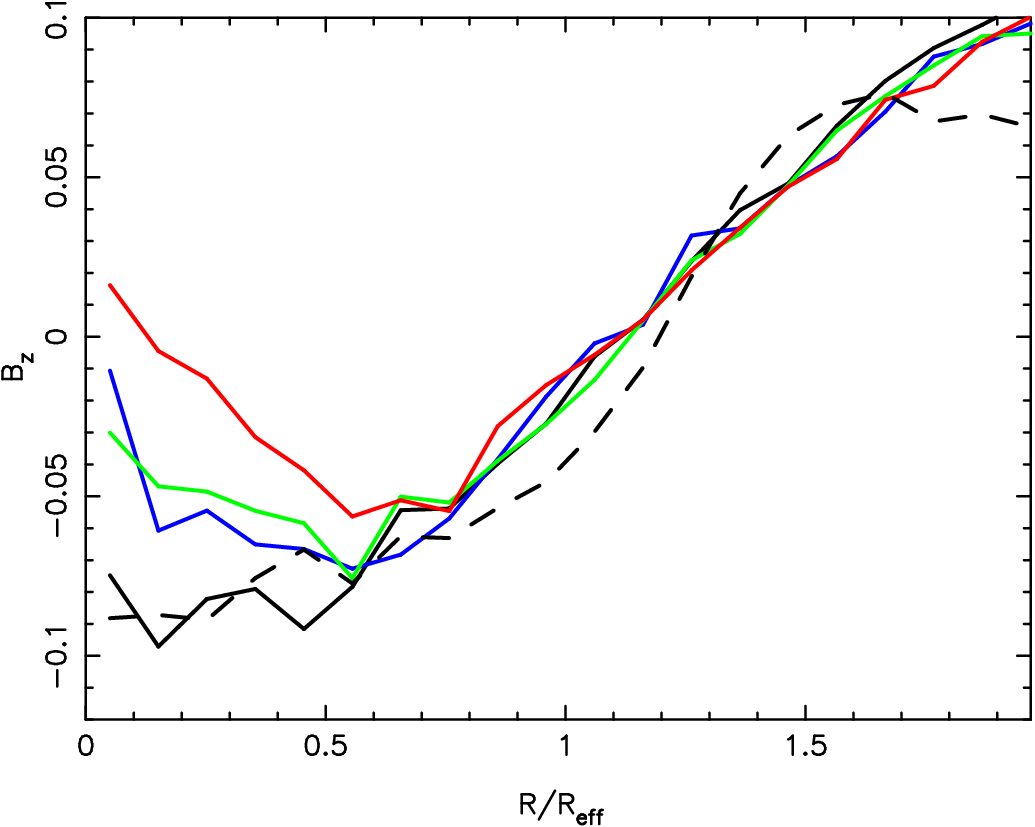}
\vskip0.2truecm
\caption[]{Profiles of 2-D (top) and 3-D (bottom) vertical anisotropy.
  The black, blue, green and red lines correspond to $\Mbh = 0, 0.8,
  1.23$, and $3.0 \times 10^5~ \Msun$, respectively.  The dashed lines
  correspond to assuming that the nuclear cluster is inclined
  at $83\degrees$ instead of being perfectly edge-on.}
\label{fig:anisotropywithbh}
\end{figure}

We therefore tested whether the assumption of a perfectly edge-on
nuclear cluster may bias the modelled vertical anisotropy to negative
values if the real inclination is somewhat smaller.  The smallest
inclination at which we were able to deproject the NCD photometry was
$83\degrees$.  Using a model deprojected at this assumed inclination
and the observed kinematics, we built {\sc nmagic} models assuming
$M/L_{\mathrm NCS}=1.5$ starting from the best-fit edge-on model M2.
The dashed black lines in Figure \ref{fig:anisotropywithbh} show that
the 2-D anisotropy, $\beta_z = 1 - \sigma_z^2/\sigma_R^2$, and 3-D
anisotropy, $B_z = 1 - 2\sigma_z^2/(\sigma_R^2 + \sigma_\phi^2)$, are
barely changed compared to the edge-on case (solid black lines) and
remain negative.  Thus a negative vertical anisotropy is not an
artifact of assuming the nuclear cluster is perfectly edge-on.

We finally explore whether the recovered $\beta_z$ changes if we
include IMBHs in the models.  In Figure \ref{fig:anisotropywithbh} we
plot both the vertical anisotropies for varying $\Mbh$.  While
increasing $\Mbh$ raises the vertical anisotropy, it still remains
negative within \re.  We conclude that the NC must be vertically
anisotropic even if a black hole were present.

\subsection{Summary}

We have built dynamical models of the NC in the nearby, edge-on
late-type galaxy NGC 4244. Using particle re-sampling, we were able to
obtain a narrow distribution of weights in our {\sc nmagic} models
allowing us to use the models as initial conditions in $N$-body
simulations. Our results can be summarised as follows:

\begin{itemize}

\item We find a mass of the spheroidal component of the NC,
  $M_\mathrm{NCS} = 1.6^{+0.5}_{-0.2} \times 10^7 ~ \Msun$ within 42.4
  pc.  The mass within 15 pc is $\sim 1.0 \times 10^7 \Msun$, in very
  good agreement with the value estimated by \citet{hart_etal_11}
  using two-integral {\sc jam} models. This mass puts the nuclear
  cluster above the \Mnc-$\mathrm{M_{Gal}}$ relation.

\item The mass of the bluer disc component of the nuclear cluster is
  less well constrained and covers the range $3.6 \times 10^5 \Msun
  \la M_\mathrm{NCD} \la 14.4 \times 10^5 \Msun$.

\item Our three-integral models are consistent with no black hole as
  well as with a black hole as massive as $4.6 \times 10^5 ~
  \Msun$.  This upper limit is larger than the one allowed by
  two-integral {\sc jam} models.

\item Simulations show that the model without a black hole is stable
  against axisymmetric perturbations.  This stability derives from the
  large Toomre-$Q$ of the system.

\item Regardless of whether a black hole is present or not, and
  of whether the nuclear cluster is perfectly edge-on or not,
  $\beta_z$ and $B_z$ are both negative.  Accretion of a star cluster
  of as little as $13\%$ by mass is enough to drive $\beta_z$ to
  positive values, regardless of the orbital geometry.  It remains
  unclear, therefore, how $\beta_z < 0$ arose.

\end{itemize}

\bigskip
\noindent

\section*{Acknowledgments}
The {\sc nmagic} models in this paper were run on Albert, the
supercomputer at the University of Malta.  The simulations were run on
Albert, on the High Performance Computer Facility at the University of
Central Lancashire and on the COSMOS Consortium supercomputer within
the DIRAC Facility jointly funded by STFC, the Large Facilities
Capital Fund of BIS.  Additional low resolution test models were run
on the old linux cluster of the dynamics group at MPE. V.P.D.
  is supported in part by STFC Consolidated grant \# ST/J001341/1.

\bibliography{n4244}

\label{lastpage} 

\end{document}